\documentclass[journal]{IEEEtran}
\IEEEoverridecommandlockouts
\usepackage{cite}
\usepackage{enumitem}
\usepackage{amsmath,amssymb,amsfonts,amsthm}
\usepackage{algorithm}
\usepackage{algpseudocode}
\usepackage{graphicx}
\usepackage{textcomp}
\usepackage{xcolor}
\usepackage{relsize}
\usepackage{comment}
\usepackage{tabularx}
\usepackage{url}
\usepackage[hidelinks]{hyperref}  
\usepackage{scalerel}
\usepackage{tikz}
\usepackage{acronym}
\usepackage{siunitx}
\usepackage{multirow}
\ifCLASSOPTIONcompsoc
    \usepackage[caption=false, font=normalsize, labelfont=sf, textfont=sf]{subfig}
\else
\usepackage[caption=false, font=normalsize]{subfig}
\fi

\def\BibTeX{{\rm B\kern-.05em{\sc i\kern-.025em b}\kern-.08em
    T\kern-.1667em\lower.7ex\hbox{E}\kern-.125emX}}
\usetikzlibrary{svg.path}
\definecolor{orcidlogocol}{HTML}{A6CE39}
\tikzset{
    orcidlogo/.pic={
        \fill[orcidlogocol] svg{M256,128c0,70.7-57.3,128-128,128C57.3,256,0,198.7,0,128C0,57.3,57.3,0,128,0C198.7,0,256,57.3,256,128z};
        \fill[white] svg{M86.3,186.2H70.9V79.1h15.4v48.4V186.2z}
        svg{M108.9,79.1h41.6c39.6,0,57,28.3,57,53.6c0,27.5-21.5,53.6-56.8,53.6h-41.8V79.1z M124.3,172.4h24.5c34.9,0,42.9-26.5,42.9-39.7c0-21.5-13.7-39.7-43.7-39.7h-23.7V172.4z}
        svg{M88.7,56.8c0,5.5-4.5,10.1-10.1,10.1c-5.6,0-10.1-4.6-10.1-10.1c0-5.6,4.5-10.1,10.1-10.1C84.2,46.7,88.7,51.3,88.7,56.8z};
    }
}

\newcommand\orcidicon[1]{\href{https://orcid.org/#1}{\mbox{\scalerel*{
                \begin{tikzpicture}[yscale=-1,transform shape]
                \pic{orcidlogo};
                \end{tikzpicture}
            }{|}}}}
\makeatletter
\def\ignorecitefornumbering#1{%
     \begingroup
         \@fileswfalse
         #1%                     % do \cite comand
    \endgroup
}
\makeatother
\newacro{IoT}[IoT]{internet-of-things}
\newacro{LEO}[LEO]{Low Earth orbit}
\newacro{GEO}[GEO]{geostationary Earth orbit}
\newacro{MEO}[MEO]{medium Earth orbit}
\newacro{LoS}[LoS]{line-of-sight}
\newacro{GS}[GS]{ground station}
\newacro{AF}[AF]{amplify-and-forward}
\newacro{MRC}[MRC]{maximal ratio combining}
\newacro{SNR}[SNR]{signal-to-noise ratio}
\newacro{SINR}[SINR]{signal-to-interference-plus-noise ratio}
\newacro{LPWA}[LPWA]{low-power wide-area}
\newacro{LoRaWAN}[LoRaWAN]{long range wide area network}
\newacro{NB-IoT}[NB-IoT]{narrowband-IoT}
\newacro{HSTN}[HSTN]{hybrid satellite-terrestrial network}
\newacro{NOMA}[NOMA]{non-orthogonal multiple access}
\newacro{SS}[SS]{single satellite}
\newacro{AWGN}[AWGN]{additive white Gaussian noise}
\newacro{PDF}[PDF]{probability density function}
\newacro{CDF}[CDF]{cumulative distribution function}
\newacro{MGF}[MGF]{moment generating function}
\newacro{SR}[SR]{shadowed-Rician}
\newacro{EIRP}[EIRP]{equivalent isotropically radiated power}
\newacro{G/T}[G/T]{antenna gain-to-noise-temperature}
\newacro{OP}[OP]{outage probability}
\newacro{DF}[DF]{decode-and-forward}
\newacro{LPWAN}[LPWAN]{low power wide area network}
\newacro{SIC}[SIC]{successive interference cancellation}
\newacro{BPP}[BPP]{binomial point process}
\newacro{CM}[CM]{capture model}
\newacro{MAC}[MAC]{medium access control}
\newacro{AWGN}[AWGN]{additive white Gaussian noise}
\newacro{DtS-IoT}[DtS-IoT]{direct-to-satellite IoT}
\newacro{CSI}[CSI]{channel state information }
\newacro{PHY}[PHY]{physical}
\newacro{BW}[BW]{bandwidth}
\newacro{3GPP}[3GPP]{third generation partnership project}
\newacro{NTN}[NTN]{non-terrestrial network}
\newacro{DL}[DL]{downlink}
\newacro{UL}[UL]{uplink}
\newacro{UE}[UE]{user equipment}
\newacro{RF}[RF]{random forest}
\newacro{ML}[ML]{machine learning}
\newacro{AQI}[AQI]{air quality index}
\newacro{RMSE}[RMSE]{root mean squared error}
\newacro{RAO}[RAO]{random access opportunities}
\newacro{TDMA}[TDMA]{time division multiple access}
\newacro{NPRACH}[NPRACH]{narrowband physical uplink shared channel}
\newacro{PM}[PM]{particulate matter}
\newacro{CPCB}[CPCB]{Central Pollution Control Board}

\theoremstyle{plain}

\begin{document}
%\linepenalty=1000
\title{Efficient Transmission Scheme for LEO Satellite-Based NB-IoT: A Data-Driven Perspective  
\\
% {\footnotesize \textsuperscript{*}Note: Sub-titles are not captured in Xplore and should not be used}
% \thanks{Identify applicable funding agency here. If none, delete this.}
}

\author{Ayush Kumar Dwivedi${\textsuperscript{\orcidicon{0000-0003-2395-6526}}}$, Houcine Chougrani${\textsuperscript{\orcidicon{0000-0001-5316-3207}}}$, Sachin Chaudhari${\textsuperscript{\orcidicon{0000-0003-1923-0925}}}$, \IEEEmembership{Senior Member, IEEE},\\ Neeraj Varshney${\textsuperscript{\orcidicon{0000-0003-0752-2103}}}$, \IEEEmembership{Senior Member, IEEE},
Symeon Chatzinotas${\textsuperscript{\orcidicon{0000-0001-5122-0001}}}$, \IEEEmembership{Fellow, IEEE}

\thanks{Ayush Kumar Dwivedi (\textit{Corresponding author}) and Sachin Chaudhari are with the Signal Processing and Communication Research Center (SPCRC) at International Institute of Information Technology, Hyderabad 500032, India (e-mail: ayush.dwivedi@research.iiit.ac.in; sachin.c@iiit.ac.in).}
\thanks{Neeraj Varshney is with the Wireless Networks Division, National Institute of Standards and Technology, Gaithersburg, MD 20899 USA (e-mail: neerajv@ieee.org).}
\thanks{Houcine Chougrani and Symeon Chatzinotas are with the Interdisciplinary Centre for Security, Reliability and Trust (SnT), University of Luxembourg, L-1855 Luxembourg, Luxembourg (e-mail: houcine.chougrani@uni.lu, symeon.chatzinotas@uni.lu)}
\vspace{-0.3cm}
}

\maketitle

\begin{abstract}
This study analyses the medium access control (MAC) layer aspects of a low-Earth-orbit (LEO) satellite-based Internet of Things (IoT) network. A transmission scheme based on change detection is proposed to accommodate more users within the network and improve energy efficiency. Machine learning (ML) algorithms are also proposed to reduce the payload size by leveraging the correlation among the sensed parameters. Real-world data from an IoT testbed deployed for a smart city application is utilised to analyse the performance regarding collision probability, effective data received and average battery lifetime. The findings reveal that the traffic pattern, post-implementation of the proposed scheme, differs from the commonly assumed Poisson traffic, thus proving the effectiveness of having IoT data from actual deployment. It is demonstrated that the transmission scheme facilitates accommodating more devices while targeting a specific collision probability. Considering the link budget for a direct access NB-IoT scenario, more data is effectively offloaded to the server within the limited visibility of LEO satellites. The average battery lifetimes are also demonstrated to increase by many folds by using the proposed access schemes and ML algorithms.
\end{abstract}

\begin{IEEEkeywords}
Energy efficiency, LEO satellites, machine learning, NB-IoT, real-world datasets, transmission reduction 
\end{IEEEkeywords}

%---------Introduction-----------
\section{Introduction}
\label{intro}
\ac{LEO} satellites have revolutionised the landscape of \ac{IoT} networks, offering ubiquitous coverage that facilitates the deployment even in previously inaccessible remote areas. The recent \ac{3GPP} Release-17 work item supporting \ac{NB-IoT} in \ac{NTN} have provided great impetus to research in this direction \cite{standard_1}. \ac{NB-IoT} is a recent cellular technology standardized by the \ac{3GPP} that aims to provide improved coverage for many low-throughput, low-cost devices with low device power consumption in delay-tolerant applications. Prospective applications include utility metering, environment monitoring, asset tracking, municipal light, and waste management, to name but a few  \cite{lin2016random}. However, many \ac{LEO} constellations targeting \ac{IoT} applications provide discontinuous coverage, introducing the challenge of limited visibility duration \cite{leo_1,leo_2}. The satellites in sparse \ac{LEO} constellations make multiple passes in a day with inter-pass duration depending upon the constellation size and orbit. Moreover, the longer propagation distance and higher path loss also lead to lower data rates in \ac{LEO} satellite-based systems than terrestrial systems. Hence, traffic/overhead reduction and energy efficiency efforts are paramount in all \ac{NTN} scenarios.

The limited visibility duration of \ac{LEO} satellites necessitates an optimized access scheme to accommodate more users within the network. Transmission reduction schemes, traditionally used as a measure to improve energy efficiency in \ac{IoT} networks, can be used along with access protocols to reduce traffic in satellite-based \ac{IoT} networks. A Shewhart-based transmission reduction scheme is a simple and efficient change-detection method well suited for \ac{IoT} applications \cite{shewhart3}. This paper proposes using the Shewhart-based access scheme to address the dual challenges of limited visibility and low data rates in addition to energy efficiency, which is even more important in remote areas where satellite access is required. The number of simultaneously transmitted devices can be reduced by strategically utilising Shewhart in limited timeframes. Thus, more devices can be accommodated within the network, mitigating the constraints of limited satellite visibility. Simultaneously, the scheme enhances data transmission efficiency by reducing the transmission of redundant data, allowing for effective communication to the server within the confines of limited bandwidth.

In the context of \ac{IoT}, the Poisson traffic model is widely adopted, including in the standardization works at \ac{3GPP} \cite{standard_poisson}, and for performance analysis of random access schemes in general \cite{collPoisson}. However, with the implementation of the Shewhart-based access scheme, the average time for a new transmission does not remain independent of the last transmission as assumed in the Poisson process. On the contrary, it depends on the nature of sensed data and factors affecting the sensing environment in the case of \ac{IoT} applications. Thus, traditional assumptions like Poisson traffic may not align with real-world scenarios while analysing the performance of Shewhart-based access schemes. Real data from \ac{IoT} deployment is necessary to emulate the traffic accurately. This paper utilizes data from an \ac{IoT} testbed deployed for air pollution monitoring to analyse the performance of the proposed schemes \cite{Rajashekar,Ayu}. It evaluates the impact of the Shewhart scheme on traffic patterns, collision probabilities, effective data received at the server, and the battery lifetimes, providing insights specific to \ac{LEO} satellite scenarios.
\subsection{Related Work}
Survey studies in \cite{survey1, survey2, ingr22, ingr23} have pointed out numerous architectural challenges inherent in satellite-based \ac{IoT} networks. These studies advocate for investigating computationally simple access schemes and exploring innovative topologies to facilitate extensive \ac{IoT} connectivity on a massive scale. A novel star-of-star topology for satellite IoT has been proposed in \cite{ayush1} and \cite{nikhil}, which leverages the benefits of mega-\ac{LEO} constellations. The performance analysis of mega-\ac{LEO} constellations has been presented in \cite{Niloofar1, Niloofar2} from a stochastic geometry perspective. A similar approach has been utilized in \cite{ayush2} to study the \ac{PHY} layer aspects of satellite-\ac{IoT} like the \ac{SIC}-based decoding and multiple satellite combining schemes.

Recently, the adaptation of \ac{LPWAN} protocols like \ac{NB-IoT} and \ac{LoRaWAN} for \ac{NTN} connectivity have gained much traction. In \cite{leo_nbiot_1}, authors proposed a novel uplink resource allocation strategy considering the dynamic nature of \ac{LEO} satellite channels, differential Doppler shifts, and increased propagation delays. The study in \cite{leo_nbiot_2} proposed a robust random access technique addressing the key challenges of long delays, significant Doppler effects, and wide beams. It also demonstrated its adaptability to various \ac{NTN} configurations outlined by \ac{3GPP} for the 5G new radio system. The adaptability of LoRa modulation technology to \ac{LEO} satellite \ac{IoT} scenarios and its various aspects, such as network architecture, access mechanisms, and bandwidth, were studied in \cite{leo_lora_1}. The potential application of chirp spread spectrum (CSS), a technology traditionally employed in radar and sonar, in the realm of \ac{LEO} satellite communication systems for low-data-rate transmission was also studied in \cite{leo_lora_2}. In \cite{leo_lora_3}, the synchronization challenges were addressed through the innovative Scheduling Algorithm for LoRa to \ac{LEO} Satellites (SALSA).  Using  \ac{TDMA} instead of traditional ALOHA-based LoRa, SALSA ensures reliable communication, mitigating issues like packet drops and collisions.

Apart from scheduling algorithms, various random access algorithms and \ac{MAC} protocols have also been studied for satellite \ac{IoT} networks. A comprehensive review of \ac{MAC} protocols for nanosatellite \ac{IoT} was presented in \cite{mac}. The assessment considered channel load, throughput, energy efficiency, and complexity factors. ALOHA-based protocols and interference cancellation-based protocols emerged with notable performance metrics, yet the study emphasized the need for improved designs that balance communication performance, energy consumption, and complexity. In \cite{ra_1}, an innovative collision detection scheme for satellite-based \ac{IoT} was proposed, emphasizing enhanced access efficiency and resource utilization. The scheme enabled rapid collision detection, and the paper provided optimal preamble selection probabilities to maximize load monitoring accuracy. The constrained application protocol (CoAP) was studied in \cite{ra_2} for reliable delivery of \ac{IoT} traffic. It established an analytical model to assess system performance under non-saturated conditions, considering \ac{MAC} parameters over a random access satellite channel with a closed-loop congestion control mechanism. Various traffic models for analysing the performance of \ac{MAC} schemes in \ac{IoT} networks are also studied in \cite{traffic_1,traffic_2,traffic_3}. The Poisson model is adopted in \cite{standard_poisson} considering the \textit{Palm-Khintchine Theorem}, which states that the superposition of asymptotically large independent renewal processes with iid inter-arrival times would lead to a Poisson process. However, as shown in \cite{traffic_1}, this assumption could lead to large errors for aggregated periodic \ac{IoT} data. Even for aperiodic event-driven \ac{IoT} applications with real-world data, the Poisson model is not followed as shown in \cite{traffic_2} and \cite{traffic_3}.

Apart from random access, various transmission reduction schemes have been studied in the literature for improving the energy efficiency of \ac{IoT} networks. \cite{shewhart2} proposed a Shewhart change detection test for a real-time smart metering system. The system transmitted local power measurements to a central processing point only when deviations exceeded a predefined threshold from the last transmitted measurement. The average time to a new transmission was derived analytically. The authors in \cite{shewhart3} introduced piggybacking and interpolation techniques to further reduce the error without increasing packet transmissions in a Shewhart-based temperature monitoring system. The battery lifetime analysis showed a significant error reduction with a marginal battery lifespan decrease. In \cite{adarsh}, the authors introduced an innovative \ac{ML}-driven approach to transmission reduction in \ac{IoT}. The study comprehensively compared supervised machine learning algorithms, considering the challenges posed by memory and computation constraints on microcontrollers. 

\begin{table*}[t!]
\renewcommand{\arraystretch}{1.3}
\centering
\caption{Comparison of work done in this paper with other papers in the literature}
\label{Lit}
\begin{tabular}{|l|c|c|c|c|c|c|c|c|c|}
\hline
% \ignorecitefornumbering{\cite{lora2,lora1}}
\textbf{Scope/Reference} & \ignorecitefornumbering{\cite{shewhart3}} & \ignorecitefornumbering{\cite{leo_nbiot_2}} & \ignorecitefornumbering{\cite{leo_lora_3}} & \ignorecitefornumbering{\cite{ra_1}} & \ignorecitefornumbering{\cite{ra_2}} & \ignorecitefornumbering{\cite{traffic_1,traffic_2,traffic_3}} & \ignorecitefornumbering{\cite{shewhart2}} & \ignorecitefornumbering{\cite{adarsh}} & This paper \\
\hline \hline
LEO satellite scenario &  & \checkmark & \checkmark & \checkmark & \checkmark &  &  &  & \checkmark \\ \hline
Transmission/scheduling scheme & \checkmark & \checkmark & \checkmark &  & \checkmark &  & \checkmark & \checkmark & \checkmark \\ \hline
ML for transmission/payload reduction &  &  &  &  &  &  &  & \checkmark & \checkmark \\ \hline
Traffic-pattern/load analysis &  &  &  & \checkmark & \checkmark & \checkmark & \checkmark &  & \checkmark \\ \hline
Collision probability analysis &  & \checkmark & \checkmark & \checkmark & \checkmark &  &  &  & \checkmark \\ \hline
Battery lifetime analysis & \checkmark &  &  &  &  &  &  &  & \checkmark \\ \hline
Real-world dataset & \checkmark &  &  &  &  & \checkmark & \checkmark & \checkmark & \checkmark \\
\hline
\end{tabular}
\end{table*}

\subsection{Contributions of the paper}
With the motivation behind leveraging \ac{LEO} satellites in \ac{IoT} networks and an aim to alleviate the limitations of visibility duration and data rates, the specific contributions of this paper are as follows:
\begin{enumerate}
    \item A Shewhart-based transmission reduction scheme for efficient access in LEO satellite-based \ac{IoT} networks is proposed, which addresses the limited visibility duration challenge specific to \ac{LEO} satellites with a sparse constellation.
    \item Data from a real-world \ac{IoT} testbed focused on monitoring air pollution is utilized to analyze the performance of the access scheme. It also provides tangible insights into observed traffic patterns, differing from the commonly assumed Poisson traffic in theoretical analysis. These insights can be helpful to emulate and optimise a real-time IoT traffic generator, closely matching the inter-arrival times of new transmissions.
    \item Collision probability for an NB-IoT scenario is analyzed for various modes of transmission utilizing Shewhart. A significant increase in the network's capacity to handle a larger user base for a target collision probability is demonstrated. This is crucial in satellite-IoT as the reduced collisions allow faster connections within the limited visibility.
    \item Effective data at the server is evaluated for the link budget of the LEO satellites, highlighting how the Shewhart-based access scheme contributes to securing higher data as the satellite's visibility duration increases.
    \item Machine learning algorithms have been proposed to reduce payload by eliminating the need to transmit parameters with strong correlation. It also reduces the dependency on variations in multiple parameters while implementing Shewhart, thus leading to higher gains in terms of transmission reduction and the number of simultaneously transmitting devices. The reduced payload size is also demonstrated to improve the energy efficiency of the nodes, increasing the average battery lifetimes by many folds. This is significant in the context of satellite-IoT, which has a constrained link budget.
\end{enumerate}
\begin{figure}[t!]
\centering
\includegraphics[width=\columnwidth, keepaspectratio]{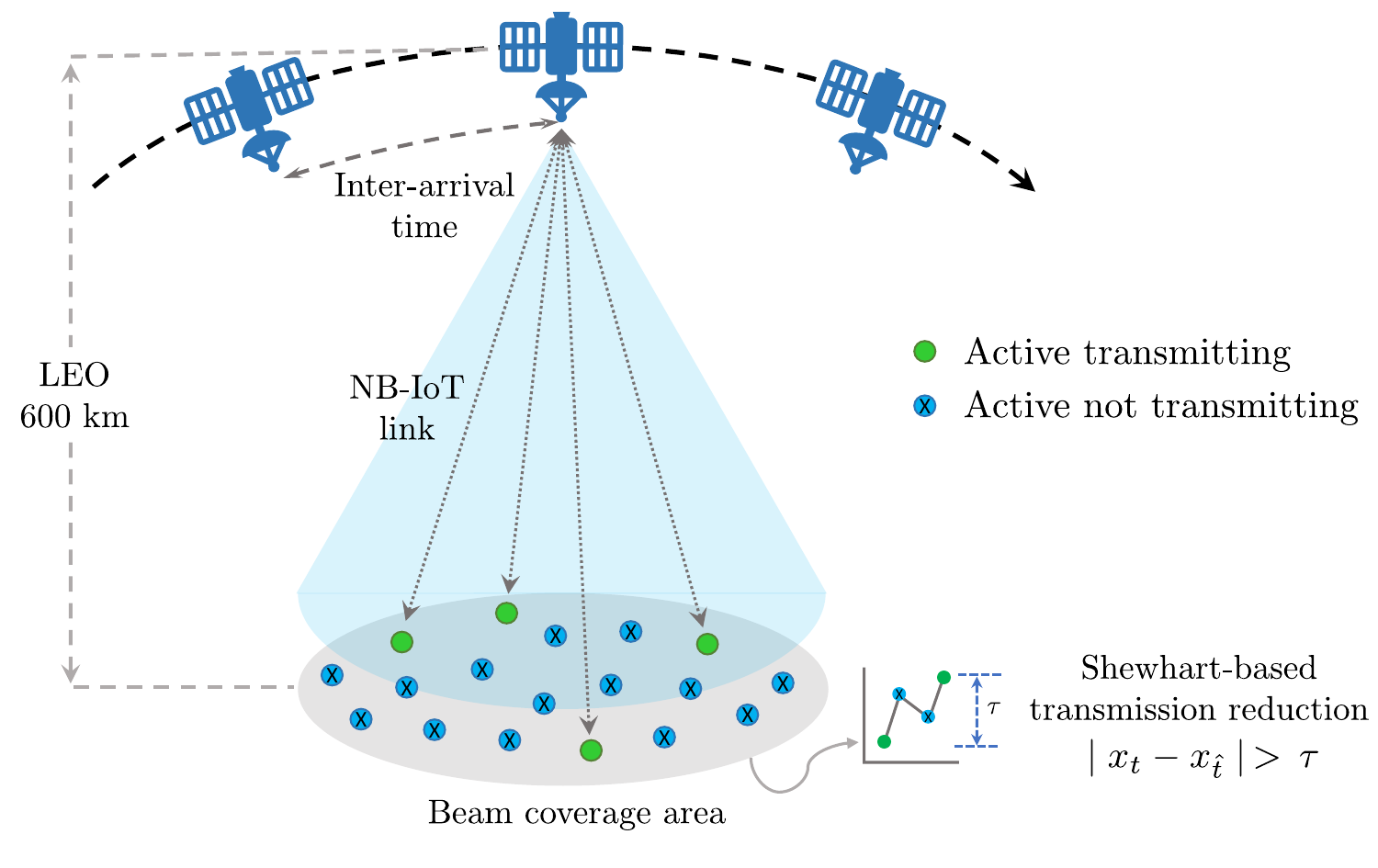}
\caption{System model: Direct-access NB-IoT network with LEO satellites at 600 km and having discontinuous coverage. All the devices under the coverage area perform Shewhart-based transmissions during the visibility period.}
\label{fig:sysModel}
% \vspace{-0.1cm}
\end{figure}

While the previous study in \cite{leo_lora_3} proposed an access scheme for \ac{LEO} satellite-based \ac{IoT} networks considering the limited visibility aspect, it utilizes \ac{TDMA} in \ac{LoRaWAN}, unlike the Shewhart-based RACH procedure being proposed in this paper for \ac{NB-IoT}. Similarly, \cite{ra_1} focused on collision detection and load estimation while \cite{ra_2} focused on designing a rate-control algorithm implemented as a CoAP congestion control algorithm. On the contrary, this paper utilises transmission reduction to reduce collisions, increase energy efficiency, and improve resource utilisation. Moreover, unlike this study, none of the above papers utilized any dataset from real-world \ac{IoT} deployment to analyse their systems. The work in \cite{shewhart2, shewhart3, adarsh} was related to Shewhart and \ac{ML}-based schemes for transmission reduction in \ac{IoT}, but none of them considered a \ac{LEO} satellite scenario. Based on the scope, Table \ref{Lit} compares the work done in this paper with other works in the literature.

The rest of the paper is organised as follows: Section \ref{sys} details the \ac{LEO} satellite system and the \ac{NB-IoT} protocol, including the link budget analysis and the method employed for calculating visibility duration and collision probability. Section \ref{subsec:dataset} outlines the measurement campaign and the steps undertaken for data preprocessing and dataset creation. The Shewhart-based transmission reduction scheme, along with the \ac{ML} algorithms and the proposed transmission modes, are presented in Section \ref{sec:schmes}. Section \ref{sec:results} then provides the outcomes and observations on various KPIs, comparing the proposed schemes with the baseline. The paper is concluded in Section \ref{sec:conclusion} with a summary of the key findings.
%
%------------System Description----------------
\section{System Description}
\label{sys}
This section presents the description of the \ac{LEO}-satellite scenario considered for this study, the associated link budget and the necessary discussion around the visibility duration of \ac{LEO} satellites.
{\renewcommand{\arraystretch}{1.2}
    \begin{table}[t!]
    \centering
    \caption{Satellite parameters for Set-1 and Set-4 configuration \cite{standard_1}}
    \label{Beam_layout}
    \resizebox{\columnwidth}{!}{%
    \begin{tabular}{|c|cc|}
    \hline
    \multirow{2}{*}{\textbf{Parameters}} & \multicolumn{2}{c|}{\textbf{LEO - 600 km}} \\ \cline{2-3} 
     & \multicolumn{1}{c|}{\textbf{Set-1}} & \textbf{Set-4} \\ \hline \hline
    Central beam center elevation angle & \multicolumn{1}{c|}{30$^\circ$} & 90$^\circ$ \\ \hline
    Central beam edge elevation angle& \multicolumn{1}{c|}{27$^\circ$} & 30$^\circ$ \\ \hline
    3 dB beam width & \multicolumn{1}{c|}{4.4127$^\circ$} & 104.7$^\circ$ \\ \hline
    Frequency band & \multicolumn{2}{c|}{S-band (i.e. 2 GHz)} \\ \hline
    Satellite EIRP density (dBW/MHz) & \multicolumn{1}{c|}{34} & 21.45 \\ \hline
    Satellite G/T (dB/K) & \multicolumn{1}{c|}{1.1} & -18.6 \\ \hline
    \end{tabular}%
    }
    \end{table}
}
\subsection{System model}\label{sys:LEO}
In this work, \ac{NB-IoT} is adopted as an \ac{IoT} communication protocol, as shown in Fig. \ref{fig:sysModel}. In the \ac{3GPP} framework, four reference scenarios have been specified depending on the orbit nature: \ac{GEO}, \ac{MEO} or \ac{LEO}, and beam type—whether steerable or fixed. All scenarios are studied under the assumptions of sub-6 GHz bands, transparent payload, and no inter-satellite link. In this work, we adopt an \ac{NTN} architecture where the \ac{IoT} devices are distributed in rural and suburban areas and connected via \ac{LEO} satellites with sparse constellation. The latter is favoured here for link budget and cost reasons. In fact, adopting a \ac{LEO} satellite will enhance the link quality due to lower distances between the ground and the orbits (compared to \ac{GEO} orbits).

To analyse the proposed algorithms, the link budget, coverage time, and achievable data rate under the proposed system model are initially evaluated. \ac{3GPP} has defined four sets of beam layout and radio frequency parameters for the payload: Set-1, Set-2, Set-3, and Set-4. Set-1 represents the best-case scenario with small spot beams. On the contrary, the Set-4 configuration represents the general and worst cases with one large beam. Table \ref{Beam_layout} provides the Set-1 and Set-4 configuration parameters for \ac{LEO} satellites at 600 km altitude.

\subsection{Link budget}\label{linkbudget}
In this subsection, the link budget for both the \ac{DL} and \ac{UL} of an \ac{LEO}-based \ac{NB-IoT} direct access system is evaluated. The receiver \ac{SNR} as a function of the \ac{EIRP}, the receiver \ac{G/T} ratio, the Boltzmann's constant $k_{\rm B} = -228.6$~dBW/K/Hz, the free-space propagation loss ${\rm PL}_{\rm FS}$, the atmospheric losses ${\rm PL}_{\rm A}$, the shadowing margin ${\rm PL}_{\rm SF}$, the scintillation loss ${\rm PL}_{\rm SL}$, and the signal bandwidth ($B$) can be calculated using
\begin{align}
    \mathrm{SNR\,[dB]} = \,&\mathrm{EIRP\,[dBW]} + G/T\mathrm{\,[dB]} - k_{\rm B}\,{\rm [dBW/K/Hz]}\nonumber\\
    & -{\rm PL}_{\rm FS}\,{\rm [dB]} - {\rm PL}_{\rm A}\,{\rm [dB]} - {\rm PL}_{\rm SF}\,{\rm [dB]}\nonumber\\
    & -{\rm PL}_{\rm SL}\,{\rm [dB]} - 10\log_{10}({B}\,{\rm [Hz]}).
    \label{eq:linkbudget}
\end{align}
An additional 3 dB loss is added to the link budget at the edge due to the assumption that the \ac{DL}/\ac{UL} \ac{EIRP} is the one at the beam centre and not at the edge of the beam.

{\renewcommand{\arraystretch}{1.4}
    \begin{table}[t!]
        \centering
        \caption{Link budget NB-IoT-to-satellite direct access}
        \label{tab:link_budget}
        \resizebox{\columnwidth}{!}{%
        \begin{tabular}{|l|cccc|}
        \hline
        \textbf{Parameters} & \multicolumn{4}{c|}{\textbf{LEO - 600 km}} \\ \hline \hline
        Transmission mode & \multicolumn{2}{c|}{Uplink (Set-1)} & \multicolumn{2}{c|}{Uplink (Set-4)} \\ \hline
        Elevation angle & \multicolumn{1}{c|}{27$^\circ$} & \multicolumn{1}{c|}{30$^\circ$} & \multicolumn{1}{c|}{30$^\circ$} & 90$^\circ$ \\ \hline
        TX EIRP {[}dBW{]} & \multicolumn{2}{c|}{-7} & \multicolumn{2}{c|}{-7} \\ \hline
        RX $G/T$ {[}dB/K{]} & \multicolumn{2}{c|}{1.1} & \multicolumn{2}{c|}{- 18.6} \\ \hline
        BW$_\text{NB-IoT}$ {[}kHz{]} & \multicolumn{2}{c|}{3.75/180} & \multicolumn{2}{c|}{3.75/180} \\ \hline
        Free space PL {[}dB{]} & \multicolumn{1}{c|}{159.7} & \multicolumn{1}{c|}{159. 1} & \multicolumn{1}{c|}{159.1} & 154.03 \\ \hline
        Atmospheric loss {[}dB{]} & \multicolumn{4}{c|}{0.1} \\ \hline
        Shadow margin {[}dB{]} & \multicolumn{4}{c|}{3} \\ \hline
        Scintillation loss {[}dB{]} & \multicolumn{4}{c|}{2.2} \\ \hline
        Polarization loss {[}dB{]} & \multicolumn{4}{c|}{3} \\ \hline
        Additional losses {[}dB{]} & \multicolumn{1}{c|}{3} & \multicolumn{1}{c|}{0} & \multicolumn{1}{c|}{3} & 0 \\ \hline
        \textbf{SNR {[}dB{]}} & \multicolumn{1}{c|}{\textbf{15.96/-0.85}} & \multicolumn{1}{c|}{\textbf{19.56/2.75}} & \multicolumn{1}{c|}{\textbf{-3.14/-19.95}} & \textbf{4.93/-11.88} \\ \hline
        \end{tabular}%
        }
    \end{table}
}

Table \ref{tab:link_budget} shows the estimated \ac{SNR} based on the Set-1 and Set-4 configuration and parameters considered in \cite{standard_1}. The evaluation considers operation in the S-band at a carrier frequency of 2 GHz. For the \ac{UL}, the \ac{SNR}s depend on the signal bandwidth. In Set-1, the best case with multiple spot beams, 15.96 dB and 19.56 dB \ac{SNR} is achieved with a 3.75 kHz single-tone configuration at the edge and centre of the beam, respectively. On the other hand, the \ac{SNR}s are reduced to -0.85 dB and 2.75 dB when using full \ac{NB-IoT} bandwidth (multi-tone) at the edge and centre of the beam, respectively. Although decent for Set-1, the obtained \ac{SNR}s for Set-4 are quite low. In set-4, the \ac{SNR}s achieved are -3.13 dB to 4.93 dB in 3.75 kHz single-tone configuration and -19.95 dB to -11.88 dB for full bandwidth multi-tone configuration. In this study, we adopt the Set-4 configuration for further analysis to show the show performance in the worst-case scenario.
%Achievable data rate
According to \cite{brandborg2022system}, the link can still be closed with the obtained \ac{SNR}s in Set-4 configuration such that one can achieve a data rate of up to 1.6 kbps for the \ac{UL}.
\subsection{Satellite visibility duration}
\begin{figure}[t!]
\centering
\includegraphics[width=0.95\columnwidth, keepaspectratio]{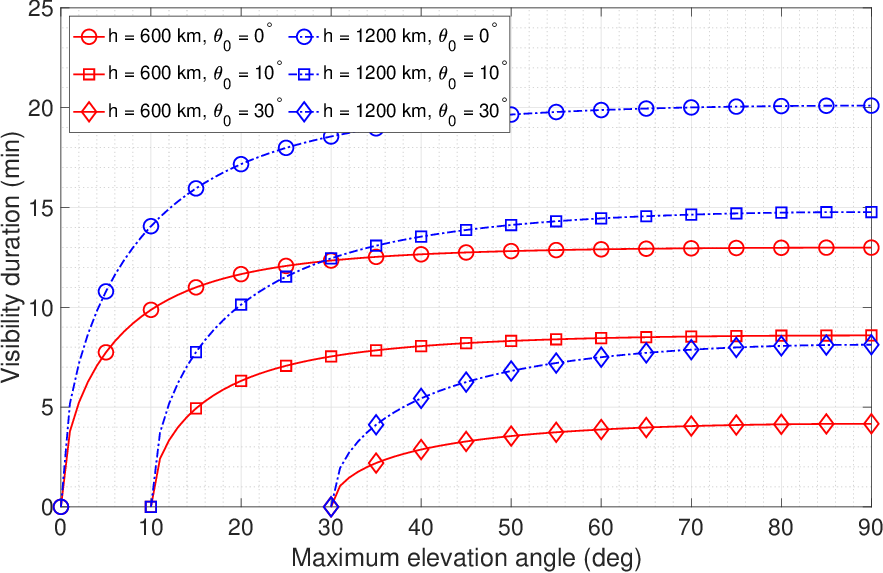}
\caption{Visibility duration $t_{\text{vis}}$ of a LEO satellite vs. the observed maximum elevation angle $\theta_{\text{max}}$ of the pass at $81^\circ$ inclination for varying altitude, $h$ and mask angle $\theta_0$. It can be observed that in a zenith pass, the satellites spend most of the visibility duration at lower elevation angles.}
\label{visibility}
\end{figure}

This study considers a \ac{LEO} constellation with discontinuous coverage. The visibility time of a \ac{LEO} satellite depends on the orbit, relative user position on the ground, and the elevation angle. A minimum elevation or mask angle is usually defined for communication with the \ac{LEO} satellites. A satellite's total visibility duration is defined as the time for which the satellite remains visible at an elevation greater than the mask angle. The visibility duration $\tau(\theta_{\text{max}})$ of the satellite at the terminal is given in \cite{doppler1} by 
\begin{align}
    & \tau(\theta_{\text{max}}) \nonumber\\
    & \approx \frac{2}{\omega_s - \omega_e \cos \theta_i} \cos^{-1} \! \left( \frac{\cos(\cos^{-1}(\frac{r_e}{r} \cos\theta_0)-\theta_0)}{\cos(\cos^{-1}(\frac{r_e}{r} \cos\theta_{\text{max}})-\theta_{\text{max}})}  \right),
\end{align}
where $r_e$ is the radius of the earth, $r$ is the satellite altitude, $\theta_i$ is the inclination of the orbit, $\theta_0$ is the angle of elevation to the satellite at a time when the satellite becomes visible to the terminal and $\theta_{\text{max}}$ is the maximum angle elevation to the satellite. Fig. \ref{visibility} shows the variation in visibility duration w.r.t maximum elevation angle at $81^\circ$ inclination for different altitudes and mask angles. The visibility duration is greater for the passes with higher maximum elevation angles. It can also be observed that in a zenith pass, the satellites spend most of the visibility duration at lower elevation angles. For example, the visibility duration for a satellite in 600 km orbit with a mask angle of $30^\circ$ is nearly 4 minutes. In contrast, the visibility duration is 13 minutes if the mask angle is considered $0^\circ$. This demonstrates that for a zenith pass, the satellite spends nearly 9 minutes to traverse from $0^\circ$ to $30^\circ$.
%
%%%%%%%%%%%%%%
%
\begin{figure*}[t!]
\centering
\subfloat[]{\includegraphics[width=0.35\linewidth, keepaspectratio]{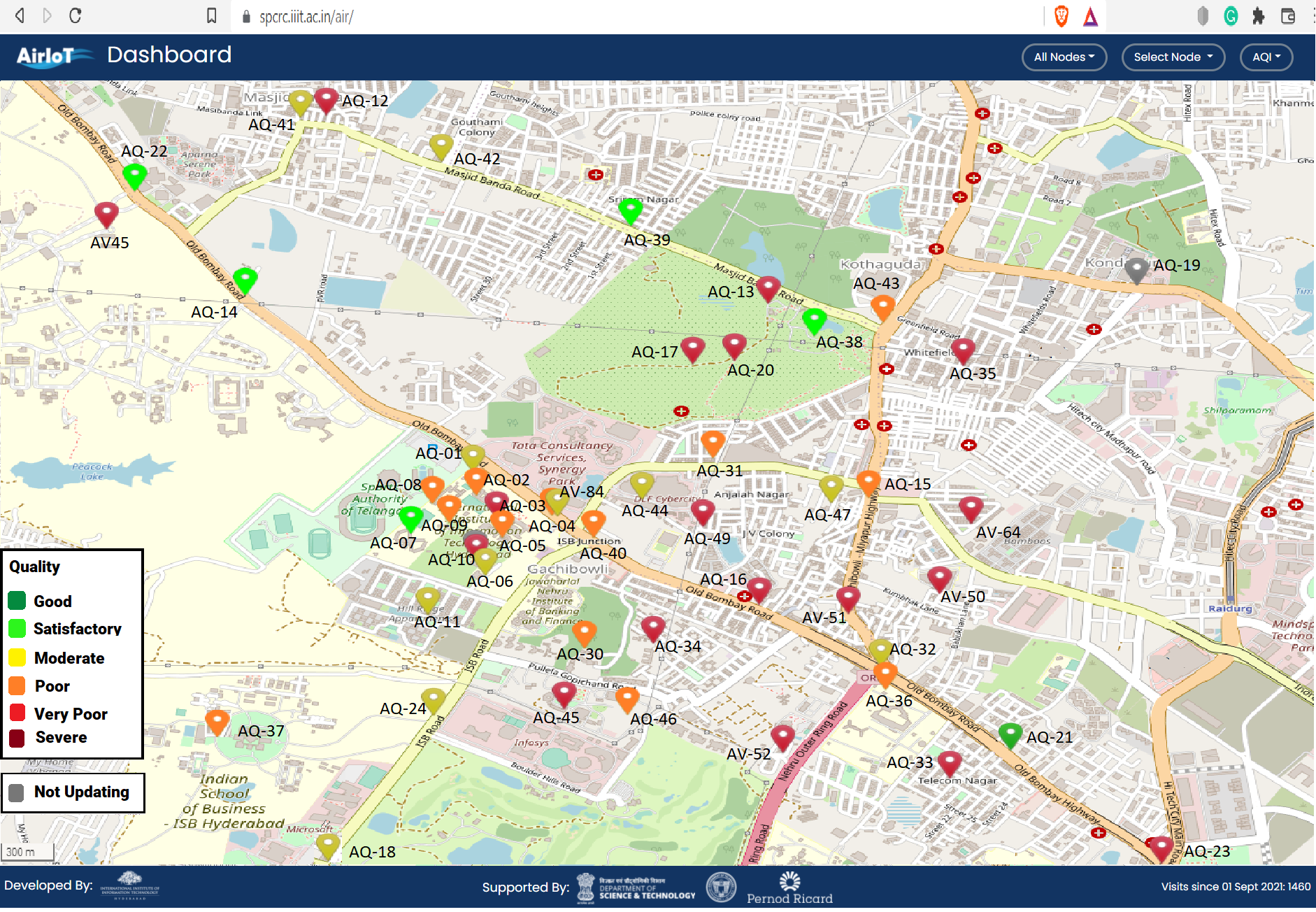}\label{fig:location}}
\hfil
\subfloat[]{\includegraphics[width=0.3\linewidth, trim={4.5cm 0 6.2cm 0},clip]{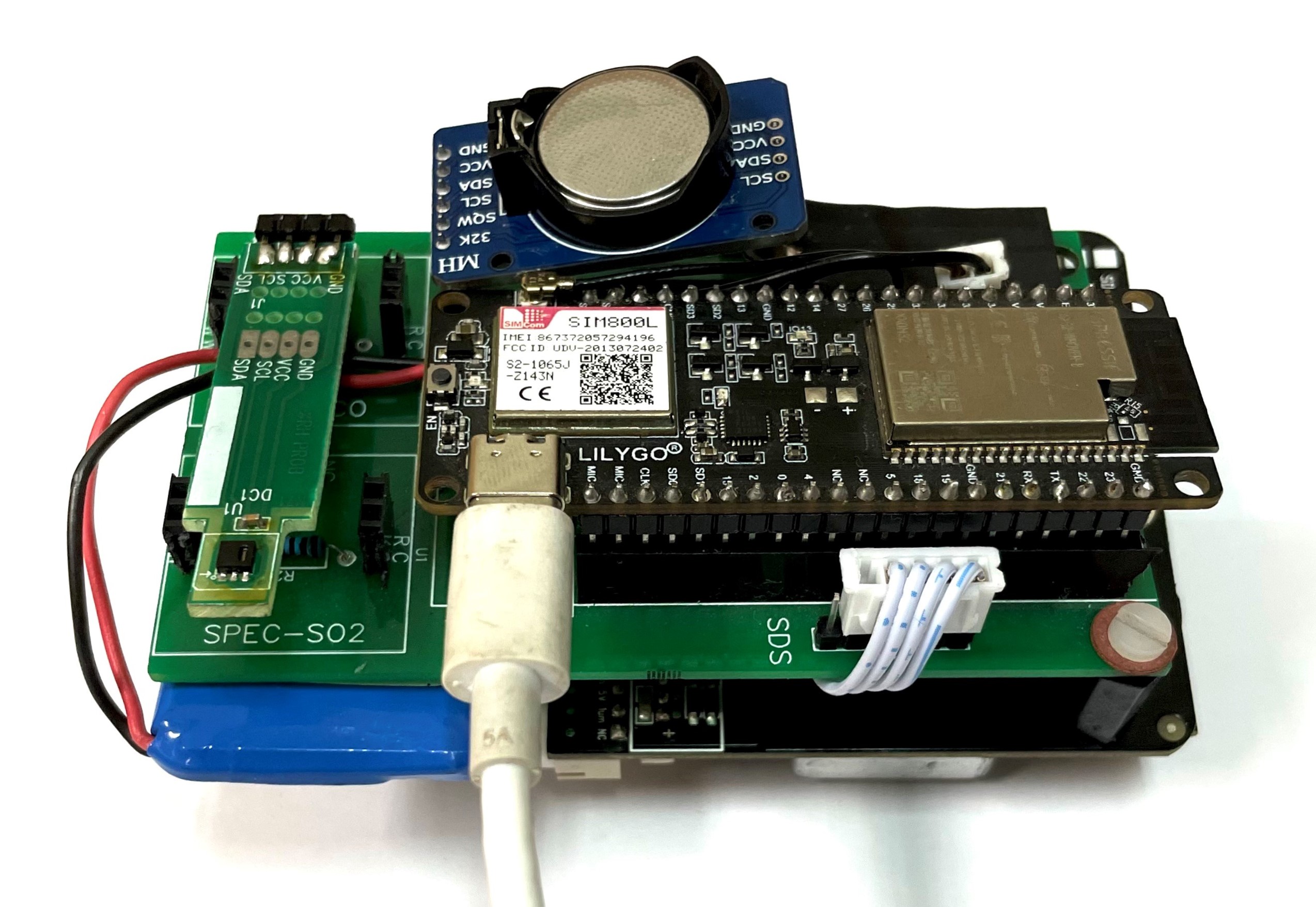}\label{fig:hardware}}
\hfil
\subfloat[]{\includegraphics[width=0.29\linewidth, keepaspectratio]{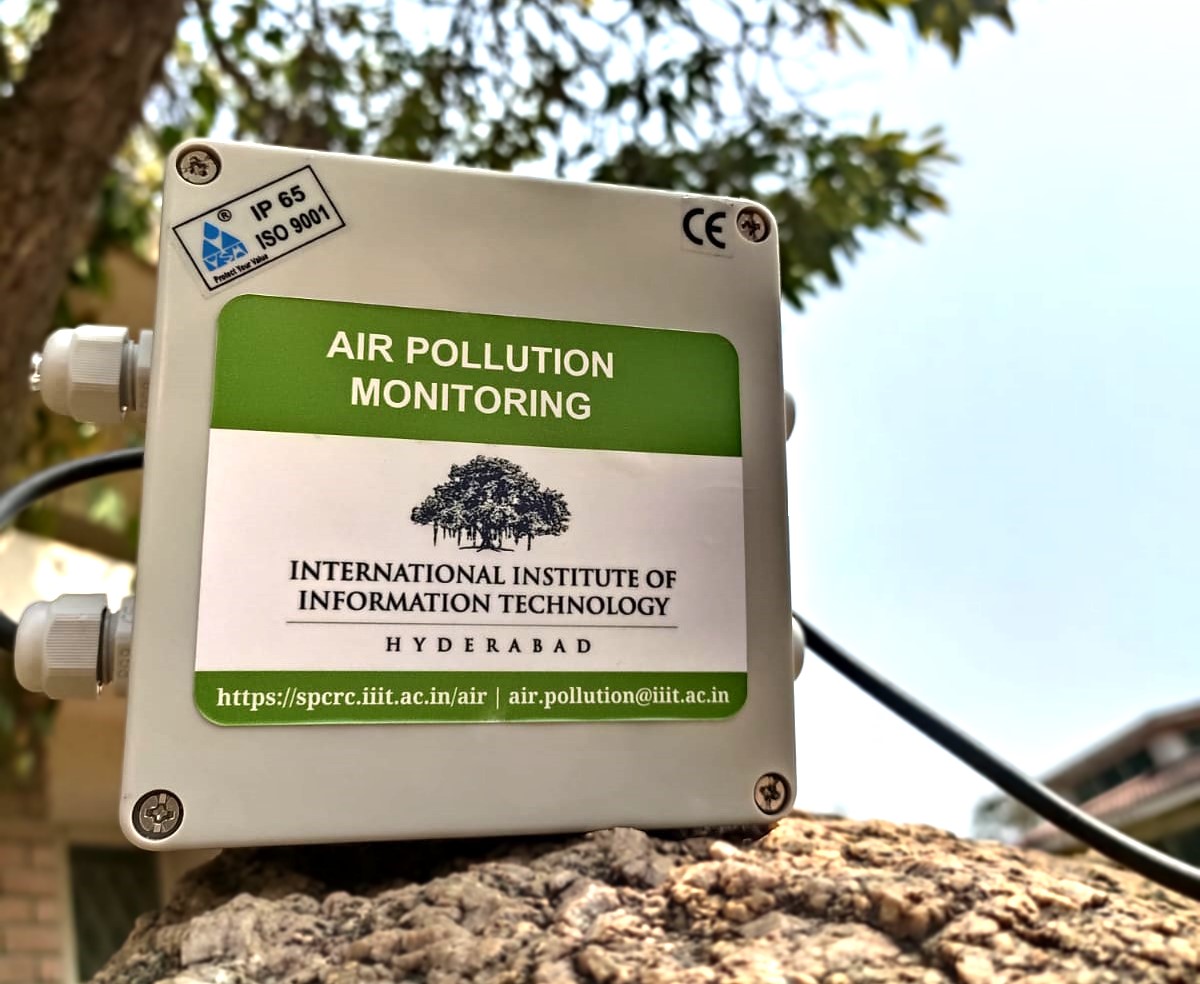}\label{fig:device}}
\caption{Deployment of air pollution monitoring devices: (a) Live data with locations on the map (b) Circuit board and sensors (c) Developed device (best viewed in colour)}
\label{deployment}
% \vspace{0.3cm}
\end{figure*}
%
%%%%%%%%%%%%%%

Although this paper considers constellation with discontinuous coverage, the observations remain relevant for constellations for continuous coverage also since it is assumed that \ac{IoT} terminals would connect and follow one satellite at a time to reduce the complexity in antenna and signal processing aspects.
%For a \ac{LEO} satellite at 600 km, a typical pass duration is about 3.3 minutes \cite{9217520}.
%
\subsection{Communication steps in NB-IoT}

In NB-IoT, communication unfolds in two distinct phases. The initial phase, known as the access or contention phase, involves nodes vying for connection establishment with the eNB through the \ac{NPRACH}. This is achieved by transmitting a preamble during \ac{RAO}, which can present 12 to 48 connection possibilities as subcarriers. Contesting nodes randomly select an index within these possibilities for preamble transmission. Collisions resulting from two or more nodes choosing the same preamble index lead to retries and failures. Successful preamble transmission initiates the data phase, exchanging messages on allocated resources without contention. Despite this, failures may still occur due to congestion or transmission errors. However, this paper focuses on the collision probability during the random access phase of \ac{NB-IoT}.

To compute the collision probability, consider an individual subcarrier in a particular \ac{RAO} with $m_{\text{RAO}}$ subcarriers \cite{collP}. Let $X_i$ be the random variable representing the number of nodes that select the $i^\text{th}$ subcarrier for transmitting their preamble in the same \ac{RAO} for a network with $N$ nodes. Then, the probability $\mathbb{P}[X_i = k]$, that $k$ nodes select the same subcarrier can be written as
\begin{align}
    \mathbb{P}[X_i = k] = \binom{N}{k} \left(\frac{1}{m_{\text{RAO}}}\right)^k \left(\frac{m_{\text{RAO}} - 1}{m_{\text{RAO}}}\right)^{N-k} . 
\end{align}
A transmission is considered successful only if a preamble is selected by one node. Therefore, the  number of nodes that reattempt the procedure can be written as
\begin{align}
    N_{\text{coll}} &= N - \left(m_{\text{RAO}} \times \mathbb{P}[X_i = 1] \right) \nonumber \\
    & = N \left(1-e^{-N/m_{\text{RAO}}}\right).
\end{align}
Each of the colliding devices selects another sub-carrier after a back-off (BO) window with probability $P_{\text{BO}}$. The collision probability for a node in an arbitrary \ac{RAO} is given in \cite{collDerivation} by
\begin{align}
    \mathcal{P}_{\text{coll}} = 1 - \exp\left(-\frac{N_{\text{coll}} \, P_{\text{BO}}}{m_{\text{RAO}}}\right).
\end{align}
\begin{figure}[t!]
\centering
\includegraphics[width=\columnwidth,keepaspectratio]{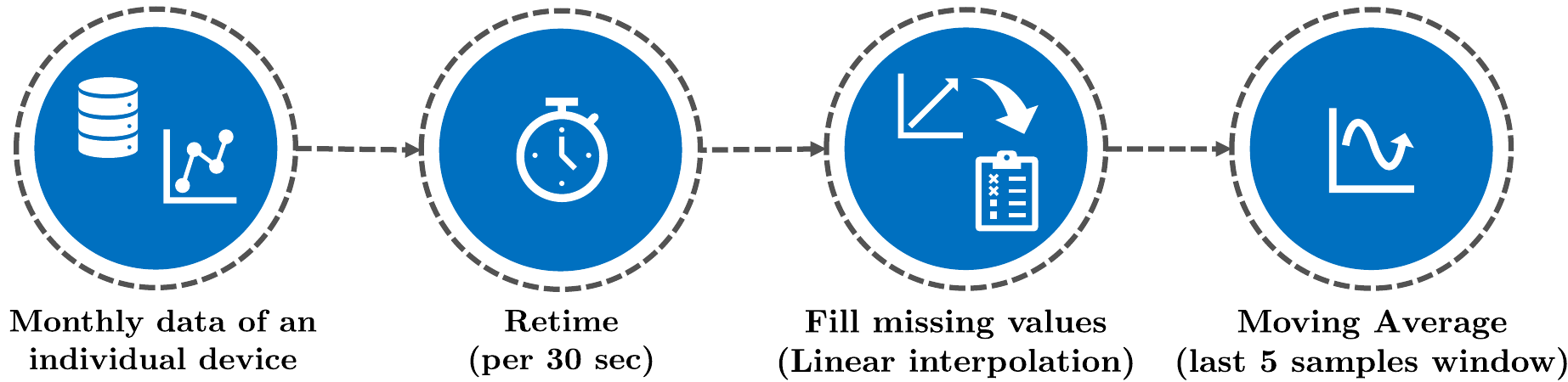}
\caption{Preprocessing steps employed on the raw sensed data to fill missing values and avoid sensor noise.}
\label{datasetProcessing}
\end{figure}
%
%
%---------------------------------%
\section{IoT testbed and dataset for experimental validation}
\label{subsec:dataset}
In this work, the performance of the proposed schemes and system model is experimentally validated using the data from an \ac{IoT} testbed, AirIoT, established by IIIT Hyderabad (IIITH), India, for air quality monitoring in a city.

\subsection{Measurement campaign}
AirIoT is a network of 49 air pollution monitoring devices deployed in an area of approximately 4 km$^2$ in the Gachibowli region of Hyderabad, the capital city of Telangana state in India. This kind of deployment could indicate many other \ac{IoT} networks, e.g. livestock, agriculture, and floating sensors for tsunami prevention, thus helping in understanding a real-life scenario. In AirIoT, the devices were strategically placed to cover urban, semi-urban and green areas, as shown in Fig. \ref{fig:location}. Every device consists of the following components: a Nova SDS011 sensor for measuring the concentration of \ac{PM} with a diameter of 2.5 and 10 microns or less (PM2.5 and PM10), a Sensirion SHT21 sensor for measuring ambient temperature and humidity, a communication module (SIM800L and eSIM), a real-time clock (RTC), and a lithium polymer (LiPo) battery. These components are connected to an ESP-32 microcontroller that senses the data every 30 seconds and offloads to ThingSpeak, a cloud-based server. The device is powered through an AC-DC power adapter and a Li-Po battery, as shown in Fig. \ref{fig:hardware}. It is enclosed in an IP-65 box with dedicated inlets to facilitate the free ambient air flow and safeguard against environmental wear and tear, as shown in Fig. \ref{fig:device}.

The entire network was deployed in a phased-wise manner starting from Dec 2019. The air quality data of the region has been collected for more than three years. A comprehensive network and dataset analysis has been presented in \cite{Rajashekar,Ayu}. A subset of the dataset collected from November 2019 to January 2023 obtained from 49 devices is used in this work.
\begin{figure*}[t!]
\centering
\includegraphics[width=\textwidth,keepaspectratio]{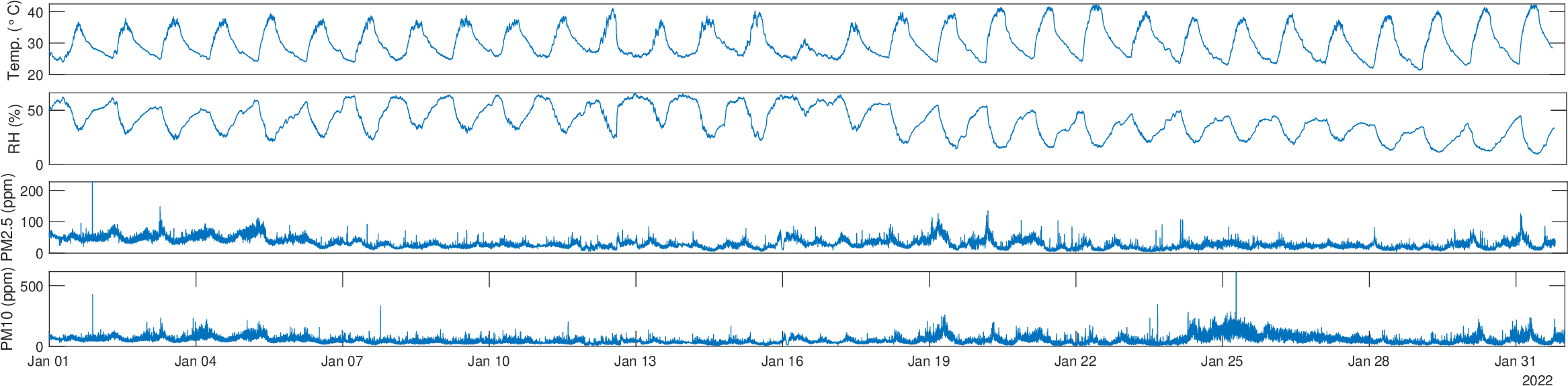}
\caption{An example snapshot of the pre-processed time-series data collected from every device. This study is based on sensing of four parameters, namely Temperature (in $^\circ$C), Relative Humidity (RH in \%), concentration of PM2.5 (in ppm) and concentration of PM10 (in ppm). It can be observed that the temperature and RH values exhibit clear periodicity, depicting the day-night cycles; however, the PM concentration is largely aperiodic, depicting its dependency on anthropogenic activities. It is also worth noticing that the temperature and RH time series hardly encounter any abrupt change in values; however, the PM time series has many abrupt changes in concentration values.}
\label{timeseries_data}
\vspace{0.3cm}
\end{figure*}
\begin{figure*}[t!]
\centering
\includegraphics[width=\textwidth,keepaspectratio]{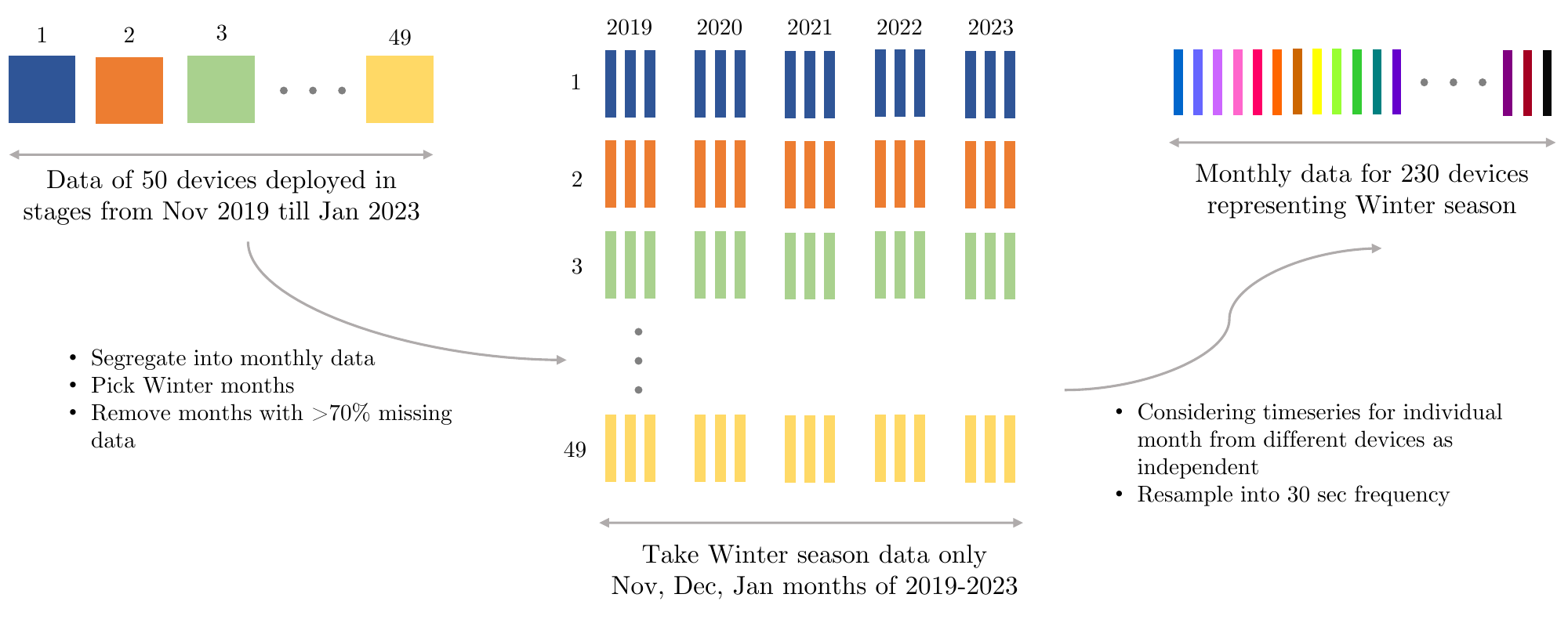}
\caption{A visual representation of the process followed to create a monthly dataset virtually representing 230 devices, as utilized for this study (best viewed in colour).}
\label{preprocessing}
\end{figure*}
\subsection{Data preprocessing}
Every device offloads a vector of data points for every sensing instance in the following format:
\begin{itemize}
    \item \textbf{device\_id}: A unique identification number specific to every device.
    \item \textbf{created\_at}: Time at which the sensor values are sensed in \texttt{"yyyy-MM-dd HH:mm:ss Z"} format.
    \item \textbf{PM10}: Concentration of PM10 in \si{\micro\gram\per\cubic\meter}.
    \item \textbf{PM2.5}: Concentration of PM2.5 in \si{\micro\gram\per\cubic\meter}.
    \item \textbf{RH}: Relative humidity values in terms of \%.
    \item \textbf{Temp}: Temperature value in \si{^\circ C}.
    \vspace{0.2cm}
\end{itemize}
The collected data was preprocessed following the steps shown in Fig. \ref{datasetProcessing}. Although the devices are programmed to sense every 30 seconds, the entire dataset was re-sampled at 30 seconds to ensure that no timestamp in the time series is missed due to data loss. The missing values were filled utilizing a simple linear interpolation technique. As a last step, a moving average filter of window size five samples (2.5 minutes) was applied to replicate the usual practice of IoT devices to avoid spurious or noisy behaviour of the sensors.  Fig. \ref{timeseries_data} shows an example snapshot of the final time series data obtained for analysing the performance of the proposed access schemes in this paper. The temperature and relative humidity (RH) values exhibit apparent periodicity, reflecting the predictable day-night cycles. In contrast, the PM concentration is largely aperiodic and lacks a distinct periodic pattern, indicating its reliance on anthropogenic activities. Also, the temperature and RH time series demonstrate minimal abrupt changes in values, maintaining a relatively steady trend. On the other hand, the PM time series displays numerous abrupt changes in concentration values, highlighting its sensitivity to dynamic and rapidly changing environmental factors.

\subsection{Dataset creation}
Fig. \ref{preprocessing} shows the process of creating the final dataset from the preprocessed data. Data from all the devices was divided into monthly chunks for each device. There have been instances of data loss primarily because the devices had to be returned to the lab for regular repair and maintenance. Additionally, the devices were brought for seasonal calibration at regular intervals and to make firmware upgrades. Hence, the data for months with more than 70\% missing values was discarded. After that, the data was arranged monthly from 2019 to 2023. The data from November to January, representing the winter months in India, was then collected for this study. Since the PM has a high seasonal correlation, the analysis in this paper has been restricted to one representative season only to avoid any seasonal bias. The variations in all four parameters sensed in this study heavily depend on local anthropogenic activities. Therefore, the time series data for individual months from different devices can be considered independent and virtually treated as data from different devices. As a result, monthly data virtually representing 230 devices was curated for this study.
%
%------------Transmission Reduction----------------
\section{Transmission reduction schemes} \label{sec:schmes}
In this work, multiple modes of operation for the \ac{IoT} devices are envisioned to reduce data transmission from the devices, thus generating varied traffic patterns. Additionally, \ac{ML} algorithms are proposed to decrease payload by excluding the transmission of strongly correlated parameters which can be predicted at the server. This approach diminishes reliance on variations in multiple parameters when implementing transmission reduction techniques, thus leading to higher gains. The proposed transmission reduction scheme and machine learning models are described below.
\subsection{Transmission reduction using Shewhart}
The Shewhart test for transmission reduction is a simple change detection algorithm \cite{shewhart1}. Its ability to detect large changes and simple implementation makes it an attractive choice for \ac{IoT} applications, as demonstrated in \cite{shewhart2} and \cite{shewhart3}. In Shewhart, newly sensed data is transmitted only if it differs from the previously transmitted value by more than a threshold. In other words, if $x_t$ is the sensed data at $t^{\text{th}}$ time instance, then a transmission is triggered if
\begin{align}
    \mid x_t - x_{\hat{t}} \mid \,>\, \tau,
\end{align}
where $x_{\hat{t}}$ is the last transmitted value and $\tau$ is the threshold. If a new value is not received at the server, the last received value is also considered to be the value at the current time instance. Therefore, the maximum error at the server is bounded by the threshold. The tolerable error usually governs the selection of threshold, transmission rate and the application at large.

\subsection{Machine learning-based prediction models for non-transmitted parameters}
Supervised \ac{ML} algorithms can eliminate the need to transmit all the sensor values or parameters among a set of strongly correlated parameters. In this work, the concentration of PM2.5 is strongly correlated with the concentration of PM10 since, by definition, the concentration of PM10 also encompasses the concentration of PM2.5 pollutants. Hence, by transmitting PM10, the transmission of PM2.5 can be avoided, thus reducing the payload size. While the sensor data in our investigation comprises only four parameters, reducing a single parameter may not intuitively lead to a significant decrease in payload size. Nonetheless, the \ac{ML} algorithms introduced in this paper for payload reduction are designed with the broader context of \ac{IoT} applications in mind, such as energy monitoring, which may involve numerous parameters. In these diverse scenarios, payload reduction holds substantial advantages for enhancing the battery lifetime of devices, as demonstrated in Section \ref{sec:batterylifetime} of this paper. The traffic generated from the dataset employed in this study serves as a representative example in this context.

Predicting the concentration of PM2.5 based on the concentration of PM10 is a regression problem. Therefore, we propose using simple machine learning algorithms like linear regression \cite{lr_bishop}, decision tree regression \cite{dt_mitchell}, and \ac{RF} \cite{rf_james} regression-based prediction of PM2.5 values using PM10 at the server. These algorithms have been selected based on their suitability for real-time deployment on cloud servers. A successful use-case of these \ac{ML} algorithms in transmission reduction was demonstrated in \cite{adarsh}.

\begin{figure*}[t]
\centering
\includegraphics[width=\textwidth]{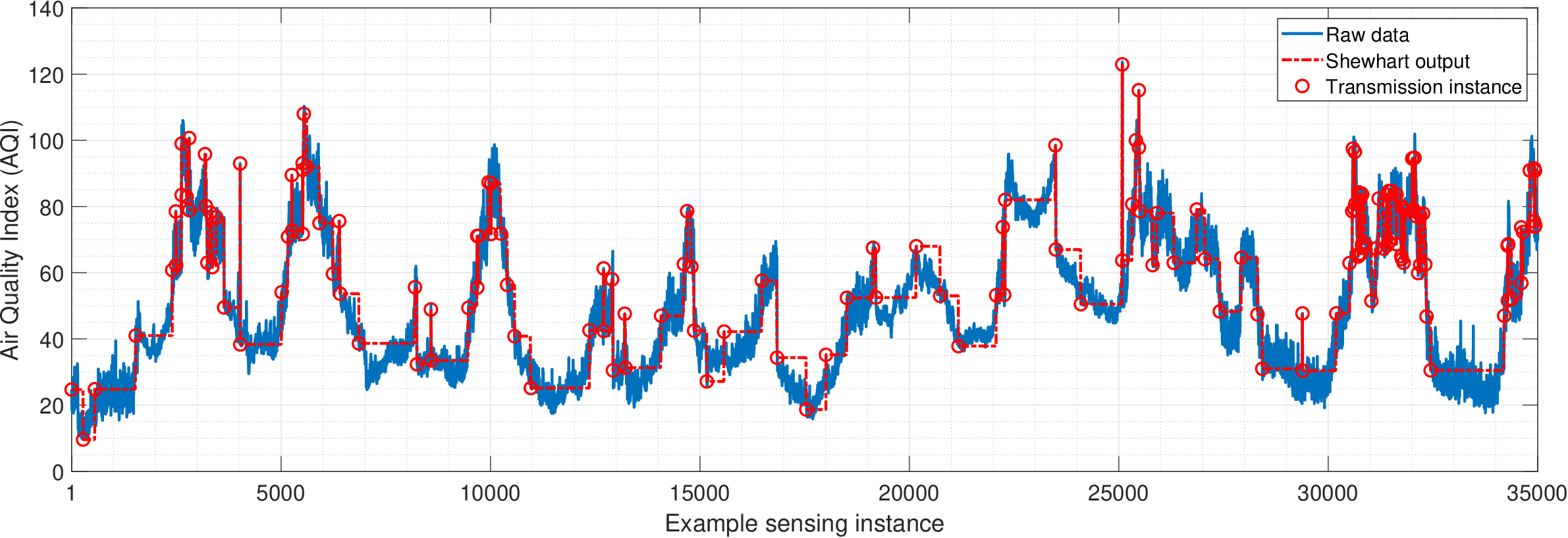}
\caption{An example snapshot of Shewhart output at the server for \ac{AQI} at a threshold of 15 index points for 100 hrs. The reduction in transmission for this snapshot is nearly 99\%.}
\label{shewhart}
\end{figure*}

\subsection{Various modes of transmission} \label{sec:modes}
In this study, Shewhart and \ac{ML}-based transmission reduction schemes have been proposed in the form of various modes of transmission. While PM2.5 and PM10 are monitored as primary pollutants in this study, the \ac{AQI} emerges as a widely recognized metric for conveying air pollution levels to the general public. Once PM2.5 and PM10 levels are known, \ac{AQI} can be easily calculated both at the node and the server using the standard formula recommended by the \ac{CPCB} in India \cite{cpcb} as shown below:
\begin{align}
    \text{AQI} & = \max\,({I_p}), \nonumber \\
    \text{where, }I_p &= \frac{I_{\text{HI}} - I_{\text{LO}}}{B_{\text{HI}} - B_{\text{LO}}} (C_p - B_{\text{LO}}) + I_{\text{LO}}, \nonumber
\end{align}
$C_p$ is the pollutant concentration, and $p = 1,2,3,\cdots,n$ is the number of pollutants. Here, $B_{\text{HI}}$ represent the breakpoint concentration greater or equal to the given concentration, $B_{\text{LO}}$ is the given concentration, $I_{\text{HI}}$ is the \ac{AQI} value corresponding to $B_{\text{HI}}$, $I_{\text{LO}}$ is the \ac{AQI} value corresponding to $B_{\text{LO}}$.

This study uses different transmission modes based on specific use cases. The list of proposed modes of transmission and the corresponding schemes are as follows:
\begin{itemize}
    \item \textbf{Mode 0}: Transmit all the parameters as soon they are sensed. \textit{This mode is equivalent to not applying any transmission reduction technique and is representative of a baseline to compare the performance of all other proposed schemes.}
    \item \textbf{Mode 1}: Apply Shewhart on all the parameters and transmit all of them if one or more parameters change significantly.
    \item \textbf{Mode 2}: Apply Shewhart on PM10 only and transmit only PM10 value to the server. When only PM10 is transmitted, the concentration of PM2.5 corresponding to every PM10 value is predicted using the \ac{ML} algorithm at the server, and the \ac{AQI} is calculated subsequently.
    \item \textbf{Mode 3}: Apply Shewhart on \ac{AQI} only, however transmit PM10 if \ac{AQI} changes significantly. The concentration of PM2.5 corresponding to every PM10 value is predicted using \ac{ML} algorithms at the server, and \ac{AQI} is calculated subsequently.
\end{itemize}
Mode 1 suits the scenarios demanding detailed data on all parameters, while for situations where only \ac{AQI} information is needed, particularly in public information systems, and higher capacity is desired, either Mode 2 or Mode 3 can be employed.
%
%------------Results----------------
\section{Results and observations}
\label{sec:results}
{\renewcommand{\arraystretch}{1.2}
\begin{table}[t]
\centering
\caption{List of parameters and their values utilized for performance analysis in this work}
\label{tab:parameters}
\resizebox{\columnwidth}{!}{%
\begin{tabular}{|l|l|}
\hline
\textbf{Parameter} & \textbf{Value} \\ \hline \hline
LEO satellite altitude & 600 km \\ \hline
Coverage type & Sparse constellation \\ \hline
System bandwidth & 180 kHz \\ \hline
Sub-carrier width & 3.75 kHz \\ \hline
No. of sub-carriers per RAO & 48 \\ \hline
RAO periodicity & 160 ms \\ \hline
UL data rate & 1.6 kbps \\ \hline
Shewhart threshold & \begin{tabular}[c]{@{}l@{}}Temp: 0.5$^\circ$C, RH: 5\%, AQI: 15\\ PM2.5: 5 ppm, PM10: 5 ppm\end{tabular} \\ \hline
Sensing rate & 30 sec \\ \hline
Back-off (BO) & $[0, 256 \cdot 2^{i}]$ sec, $i=10$ \\ \hline
Beam visibility window & 246.9 sec (set-4) \\ \hline
\end{tabular}%
}
\end{table}
}
{\renewcommand{\arraystretch}{1.3}
\begin{table*}[t!]
\centering
\caption{Performance of various modes of operation in terms of the number of simultaneously transmitting nodes, \% reduction and RMSE}
\label{tab:shewhartPerformance}
\begin{tabular}{|cc|c|c|cc|}
\hline
\multicolumn{2}{|c|}{\textbf{Transmission Mode}} &
  \textbf{\begin{tabular}[c]{@{}c@{}}Simultaneously\\ Tx. Nodes\end{tabular}} &
  \textbf{\begin{tabular}[c]{@{}c@{}}\%\\ Reduction\end{tabular}} &
  \multicolumn{2}{c|}{\textbf{RMSE}} \\ \hline \hline
\multicolumn{1}{|c|}{\textbf{M0}} & Baseline: Transmit all the parameters                      & 230 & 0     & \multicolumn{2}{c|}{0}                      \\ \hline
\multicolumn{1}{|c|}{\multirow{2}{*}{\textbf{M1}}} &
  \multirow{2}{*}{Shewhart on all parameters} &
  \multirow{2}{*}{30} &
  \multirow{2}{*}{87.16} &
  \multicolumn{1}{c|}{T: 0.16} &
  RH: 0.62 \\ \cline{5-6} 
\multicolumn{1}{|c|}{}            &                                                                  &    &       & \multicolumn{1}{c|}{PM2.5: 1.29}  & PM10: 2.10 \\ \hline
\multicolumn{1}{|c|}{\textbf{M2}} & Shewhart on PM10 only (transmit PM10 and predict PM2.5) & 26  & 88.33 & \multicolumn{1}{c|}{PM2.5: 2.34} & PM10: 2.22 \\ \hline
\multicolumn{1}{|c|}{\textbf{M3}} &
  \begin{tabular}[c]{@{}c@{}}Shewhart on AQI only (transmit PM10 and predict PM2.5,\\ calculate AQI at the server)\end{tabular} &
  8 &
  96.54 &
  \multicolumn{1}{c|}{PM2.5: 4.23} &
  PM10: 7.56 \\ \hline
\end{tabular}
\end{table*}}
In this section, we present the performance of the proposed access schemes with various transmission modes and compare them with the baseline method through different performance metrics. These metrics include the percentage reduction in transmission w.r.t the threshold in Shewhart, \ac{RMSE} of the sensed parameters at the server, the average number of simultaneously transmitting nodes, collision probability during the access/contention phase of \ac{NB-IoT}, the effective data received at the server within the visibility duration, and the expected battery lifetime. The \ac{RMSE} and the reduction in transmissions represent the efficiency of the proposed scheme, while the other metrics represent its suitability for the \ac{LEO} satellite scenario. Unless stated otherwise, all the system-level simulations have been performed using the parameters shown in Table \ref{tab:parameters}. The threshold values for Shewhart are carefully chosen, considering the sensor's accuracy tolerance and the nominal values of the sensed parameter in the specific environment. For instance, the Sensirion SHT21 sensor, employed for temperature and humidity sensing, exhibits an accuracy tolerance of $\pm 0.3^\circ C$ and $\pm 2\%$, respectively. Since the temperature in Hyderabad, India, typically ranges between 10-20°C in winter, the thresholds for temperature and humidity are set at $0.5^\circ C$ and 5\%, respectively. Similarly, the Nova SDS011 sensor used for PM monitoring has a maximum accuracy tolerance of $\pm 15\%$ up to 10 ppm. Usually, PM2.5 ranges between 80 and 120 ppm, and PM10 ranges between 180 and 250 ppm during winter in Hyderabad. However, according to \ac{CPCB}, 30 ppm for PM2.5 and 50 ppm for PM10 is categorised as "Good" \cite{cpcb}. Hence, a threshold of 5 ppm is selected for both PM2.5 and PM10 values to ensure that both the trend is captured and the data is reliable, even for the "Good" category with the least concentrations. Based on similar reasoning, given that an \ac{AQI} between 50-100 is deemed satisfactory, a threshold of 15 index points is selected for \ac{AQI}.
\subsection{Performance of Shewhart in transmission reduction}
Initially, to demonstrate the working of Shewhart, an example implementation on a snapshot of a single time series for \ac{AQI} is shown in Fig. \ref{shewhart}. The snapshot presented has a duration of 100 hours with a threshold of 15 index points. The reduction in transmission instances for the presented snapshot is nearly 99\%, affirming the Shewhart scheme's efficacy. Notably, the plot illustrates a significant decrease in transmitted instances while accurately capturing the trendline and essential variations in \ac{AQI}. This highlights the substantial benefits of integrating Shewhart for transmission reduction in this study.

Table \ref{tab:shewhartPerformance} provides a comprehensive overview of the Shewhart scheme's performance across various modes of operation applied to the entire dataset. The results are evaluated in terms of transmission reduction, the number of simultaneously transmitting nodes, and the associated \ac{RMSE} in the sensor values calculated at the server. It can be observed that all three modes with the proposed Shewhart-based access scheme produce a significant reduction in transmission. In Mode 1 and Mode 2, nearly 87-88\% of transmissions are reduced, while in Mode 3, nearly 96\% of transmissions are reduced. Notably, in Mode 1, the reduction percentage is less pronounced than in Mode 2 and Mode 3 due to the application of the Shewhart scheme to all four simultaneously varying time series. Even a sudden change in any parameter can trigger a new transmission. It can also be observed that Mode 2 also demonstrates a similar reduction percentage. This can be attributed to its tracking of PM10, where spurious changes are more prominent compared to the relatively smooth temperature and humidity data. It is noteworthy and encouraging to observe that an outstanding reduction in transmission is achieved with minimal error in the received information at the server. The \ac{RMSE} values of 0.16$^\circ$C for temperature and 0.62\% for RH are negligible compared to both the nominal values of these parameters and the selected thresholds. This favourable trend extends to PM2.5 and PM10, with \ac{RMSE} values of 2.34 and 2.22 ppm, respectively. These values are also negligible compared to the nominal values and do not significantly impact the calculation of the \ac{AQI}, which is the ultimate marker of air pollution in general applications.
\begin{figure}[t]
\centering
\includegraphics[width=0.93\columnwidth]{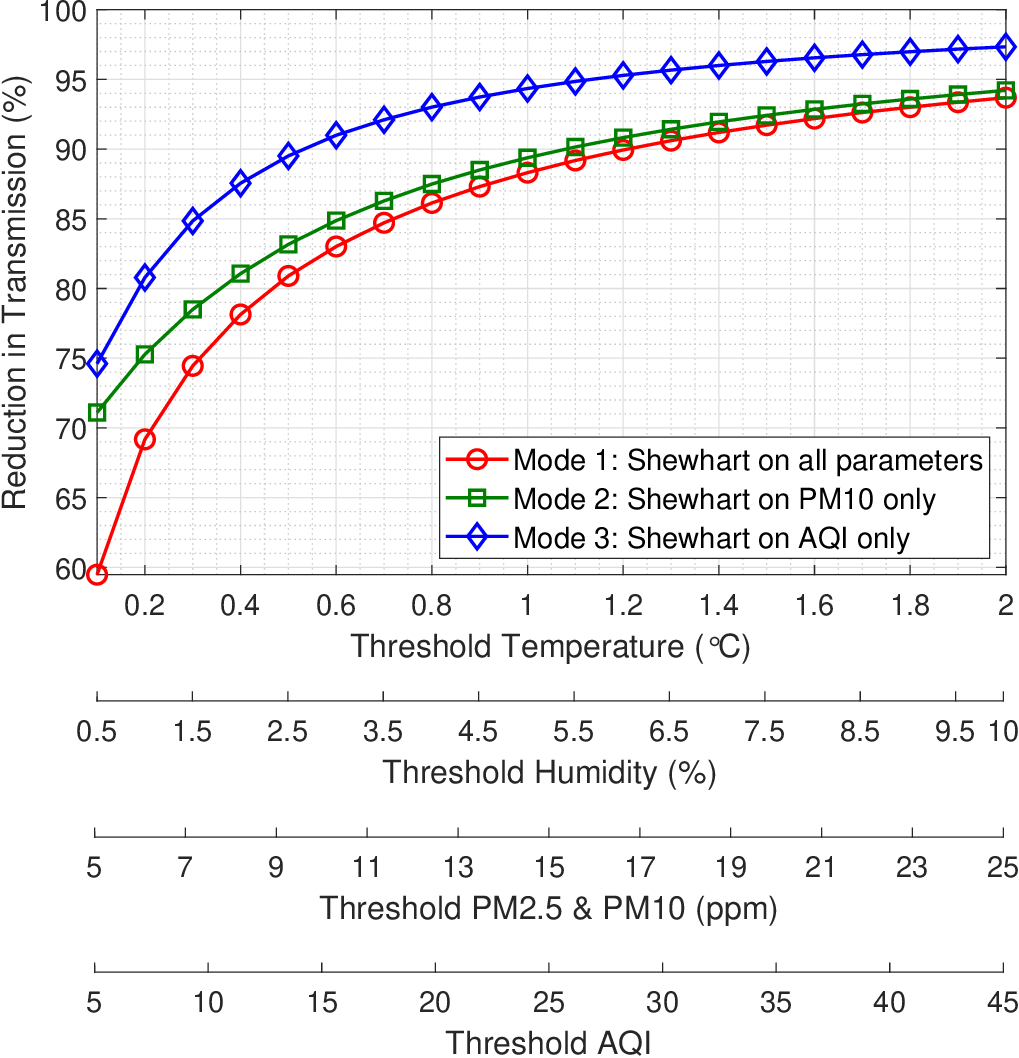}
\caption{Average percentage reduction in transmission across the entire dataset as a function of varying thresholds for different transmission modes.}
\label{fig:reductionVsThreshold}
\end{figure}

In Mode 3, a remarkable 96.54\% reduction in transmissions and a decrease in simultaneously transmitting nodes from 230 to 8 are achieved. Shewhart is applied on \ac{AQI} in Mode 3, which, being a derived quantity, exhibits stability against minute fluctuations in PM values. This stabilizing effect is explained by the \ac{AQI} calculations, as discussed in Section \ref{sec:modes}, where the \ac{AQI} primarily depends on the breakpoints of PM concentration values and not on the specific value. The breakpoint concentrations for PM10 are Good (50), Satisfactory (100), Moderate (250), Poor (350), Very Poor (430) and Severe (430+). Similarly, the breakpoint concentrations for PM2.5 are Good (30), Satisfactory (60), Moderate (90), Poor (120), Very Poor (250) and Severe (250+). Hence, any fluctuations within these breakpoints do not change the \ac{AQI}, making it a less fluctuating parameter. It can also be observed that the improved reduction in transmissions incurs only a nominal cost in terms of an increased \ac{RMSE}. For Mode 3, even though \ac{AQI} is tracked for Shewhart, PM10 values are transmitted, and \ac{AQI} is calculated at the server. Hence, the \ac{RMSE} is calculated for PM2.5 and PM10 to analyse the performance. In mode 3, the \ac{RMSE} values of PM2.5 and PM10 are 2.43 ppm and 7.56 ppm, respectively, which is considerably lower than the nominal values observed during the winter season.

In Fig. \ref{fig:reductionVsThreshold}, the percentage reduction in transmission is illustrated for different thresholds across all proposed transmission modes. Notably, the reduction percentage sharply increases with higher thresholds, a pattern that aligns with expectations. Additionally, a comparison reveals that Mode 3 exhibits a relatively higher reduction percentage than Mode 1 and Mode 2, consistent with the previously explained reasons. Here, it is worth discussing the impact of a 96\% reduction in transmission – it means sending only 4 out of every 100 newly sensed samples instead of sending all of them otherwise. As highlighted in later KPIs, this reduction allows efficient use of resources, like bandwidth and limited access time during satellite visibility, for example, scheduling 96 more devices. If the Shewhart is applied to the entire network, scheduling 96 more devices would actually mean adding the capability to accommodate nearly 100 times more devices in the network with the same amount of resources. When viewed in conjunction with the \ac{RMSE} reported above, this illustrates a significant boost in efficiency and capacity.
\begin{figure}[t!]
\centering
\includegraphics[width=0.95\columnwidth, keepaspectratio]{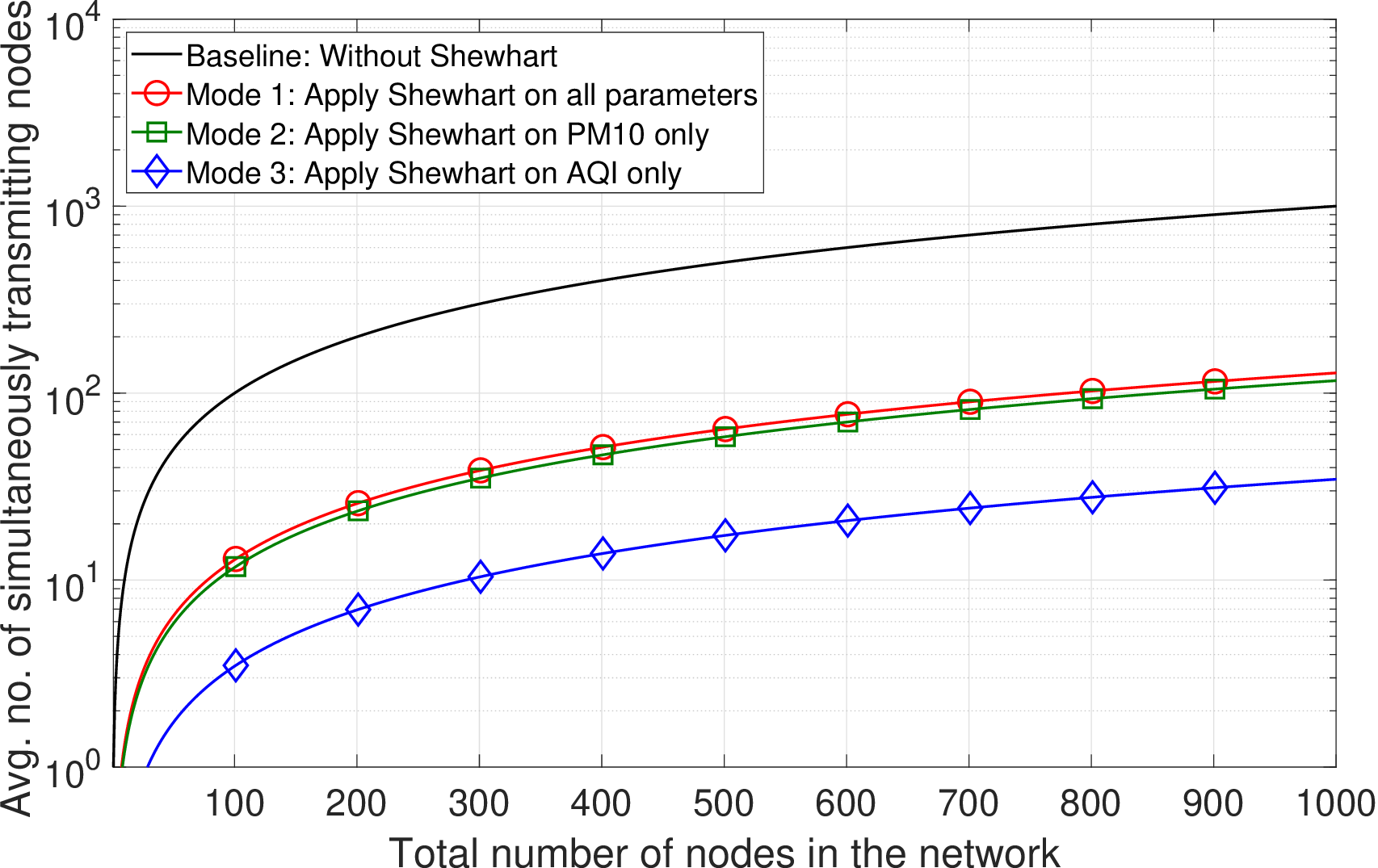}
\caption{Comparison of avg. no. of simultaneously transmitting nodes relative to the total network size for different Shewhart transmission modes. A significant reduction of the order of $10^3$ is achieved in Mode 3, while Modes 1 and 2 display a reduction of about ten times.}
\label{fig:collVsTotalNodes}
\end{figure}
\subsection{Reduction in number of simultaneously transmitting nodes}
As the network size grows, Fig. \ref{fig:collVsTotalNodes} displays the count of devices transmitting simultaneously using Shewhart. The results are extrapolated from the dataset using curve fitting to understand trends for larger networks. Notably, the number of simultaneously transmitting devices increases linearly with network size but significantly fewer in number with Shewhart when compared to the baseline scheme. For Mode 1 and Mode 2, this count is nearly ten times less than the actual devices in the network, equating to 10 out of 100 transmitting simultaneously. Mode 1 and Mode 2 exhibit similar performance, as explained earlier. Additionally, the reduction in simultaneously transmitting nodes follows an order of $10^2$, i.e., 10 out of 1000 transmitting simultaneously for Mode 3.  The result from this plot is crucial for examining other KPIs in larger networks. It serves as a tool to simulate the effects of the proposed transmission modes.
\begin{figure}[t!]
\centering
\includegraphics[width=0.95\columnwidth, keepaspectratio]{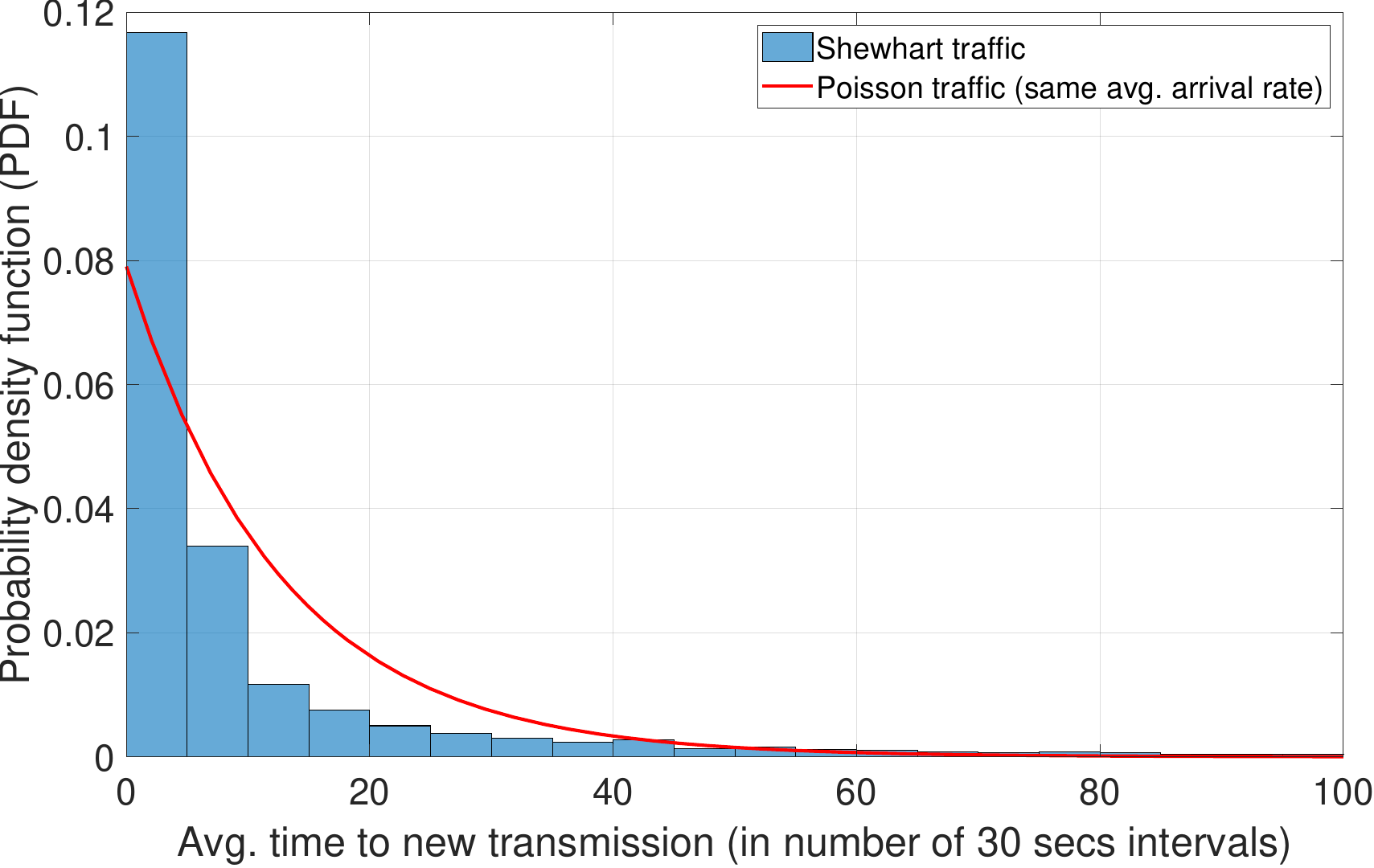}
\caption{Comparison of probability density function between Shewhart-generated traffic and Poisson arrival assumption for the average time difference between successive transmissions. It can be observed that the Poisson assumption is incompatible with the Shewhart scheme, emphasizing the value of utilizing a real dataset.}
\label{fig:traffic}
\end{figure}
\subsection{Traffic pattern generated by Shewhart}
In the context of random access systems, traditional assumptions consider the arrival rate of new packets to follow a Poisson distribution, where the difference between subsequent arrivals is exponentially distributed. However, as evidenced in Fig. \ref{fig:traffic}, the actual distribution of the average time to the new transmission, based on the real dataset used (i.e. with Shewhart), deviates significantly from Poisson traffic generated to match the average arrival rate. This emphasizes the utility of having a real dataset to generate traffic from a Shewhart-based access scheme rather than assuming Poisson arrival for system-level analysis.

The Poisson arrival process assumes that the inter-arrival time is independent of the last arrival. However, the traffic generated by Shewhart is contingent on factors such as the selection of thresholds and other parameters influencing the time series (environmental in this case). Thus, the arrival event and the inter-arrival times depend on the last arrival. Moreover, the Poisson model assumes that the number of sources is infinite, whereas in the Shewhart-based access scheme, the number of simultaneously transmitting sources is reduced drastically.
\begin{figure}[t!]
\centering
\includegraphics[width=0.95\columnwidth, keepaspectratio]{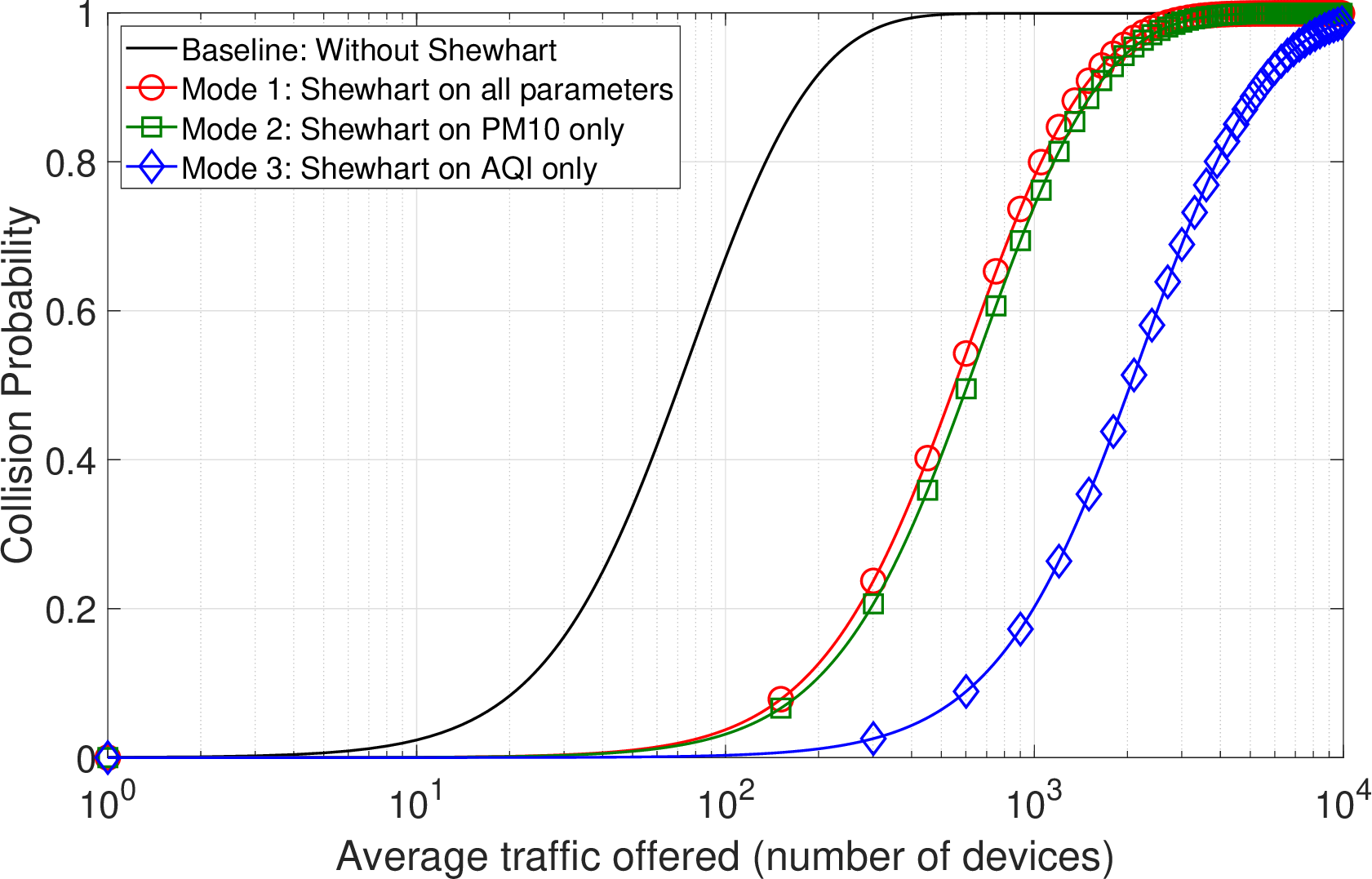}
\caption{Collision probability versus average traffic offered in terms of simultaneously transmitting nodes in the access phase of NB-IoT for RAO with 160 ms periodicity and BO index of 10. The figure illustrates a substantial increase in the achieved load by implementing the proposed access scheme for a target collision probability.}
\label{fig:collProbabililty}
\end{figure}
\subsection{Collision probability w.r.t offered traffic load}
In Fig. \ref{fig:collProbabililty}, the collision probability is plotted against the average offered traffic, measured as simultaneously transmitting nodes during the access phase of \ac{NB-IoT} for RAO with 160 ms periodicity and BO index of 10. The figure underscores the potential for increased load capacity by implementing the proposed access scheme to achieve a target collision probability. Notably, for a target collision probability of 0.4, while the baseline method supports only 57 devices, Mode 1 and 2 can accommodate nearly 450 devices, and Mode 3 can support close to 1650 devices. This is significant in the context of \ac{LEO} satellite constellation, which allows handling a large number of users owing to large beam sizes. Moreover, the reduced collisions allow faster connections, which are crucial in the context of limited visibility. This also comes with advantages like reduced power consumption (due to fewer connection re-attempts) and lower congestion in the network.

% Another noteworthy observation is the difference between the Poisson assumption estimates and real data, especially in Mode 1 and Mode 2, where the Poisson assumption falls short by a considerable margin, estimating the supported load to just above 300 devices. However, for Mode 3, the Poisson assumption aligns more closely with real data, suggesting that the reduced triggering factors for transmissions in Mode 3 make a transmission event independent of the time since the last transmission. Hence, the traffic generated by the Shewhart schemes potentially converges to a Poisson arrival process. 

%
\begin{figure}[t!]
\centering
\includegraphics[width=0.95\columnwidth, keepaspectratio]{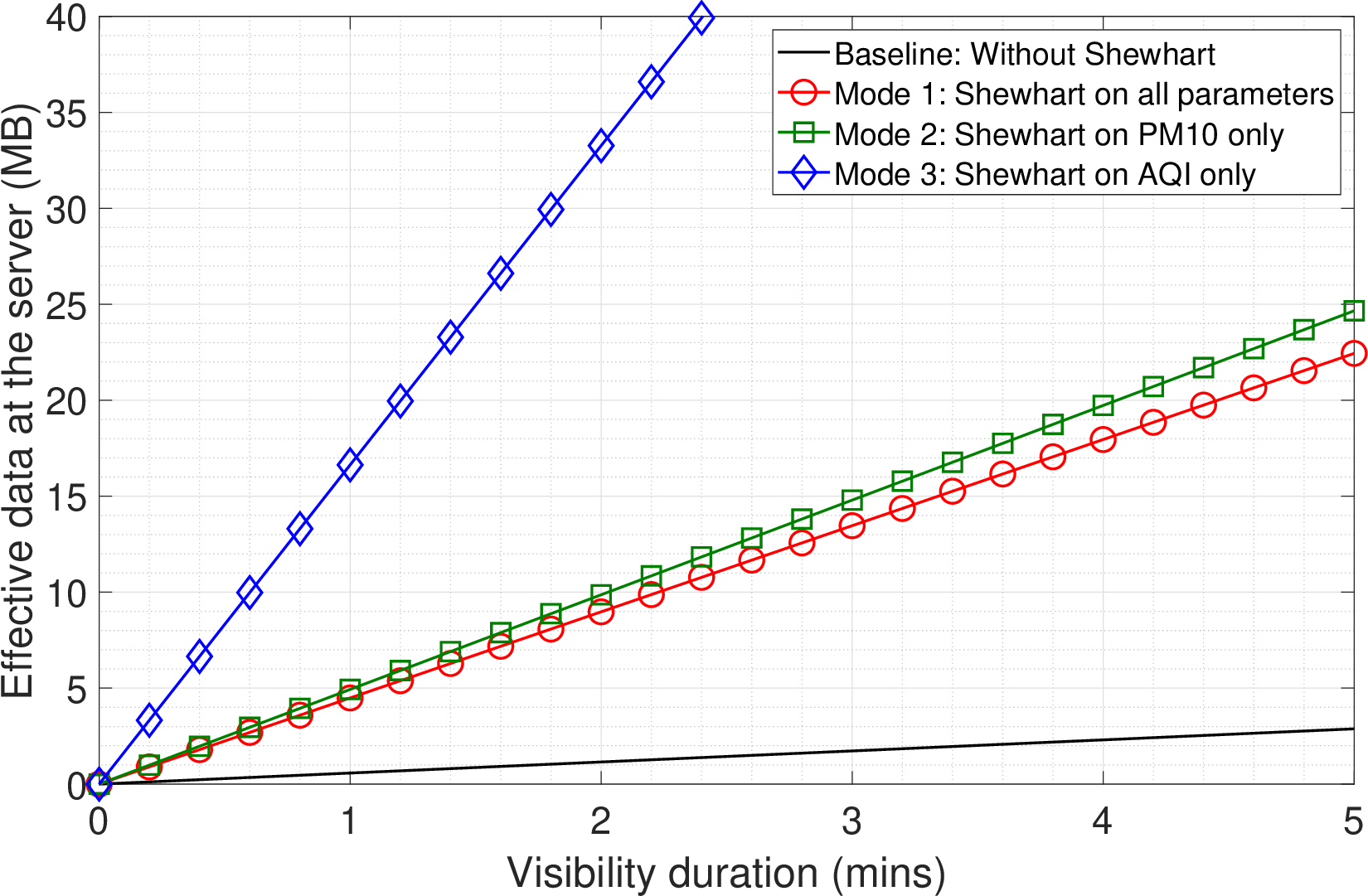}
\caption{Effective data received at the server vs. the visibility duration of LEO satellite (in the data phase) for 180 kHz bandwidth, 3.75 kHz subcarrier, and 1.6 kbps data rate. Effective data encompasses information predicted at the server, including data not transmitted by the nodes due to Shewhart.}
\label{fig:dataVsVisibility}
\end{figure}
\subsection{Effective data at the server w.r.t visibility duration}
In Fig. \ref{fig:dataVsVisibility}, the effective data received at the server is plotted against the increasing visibility duration of the \ac{LEO} satellite in the data phase of \ac{NB-IoT} communication. The plots are generated considering the link budget analysis presented in Section \ref{linkbudget} for a bandwidth of 180 kHz, a subcarrier of 3.75 kHz, and a data rate of 1.6 kbps (representing the worst-case as per Set-4 configuration). The effective data is defined to encompass the data not transmitted by the nodes due to Shewhart but predicted at the server (i.e. considering the last received value to be the sensed value at the current instance). The results showcase that the Shewhart-based access schemes optimise the limited visibility duration of \ac{LEO} satellites by effectively communicating more information using the same number of transmissions as the baseline scheme. For example, in one minute, while the baseline scheme can communicate only 576~kB of data, Mode 1 and 2 can effectively offload 5~MB, and Mode 3 can offload 16~MB data to the server. Alternatively, the gain can also be viewed in terms of the reduced time required for offloading a target amount of data. For example, what Shewhart-based access modes can transmit in less than a minute would otherwise take 5 minutes in the baseline transmission method. This significant gain is valuable, particularly in \ac{LEO} satellite-based scenarios with discontinuous coverage, where visibility duration is a crucial and limited resource. Even in scenarios with continuous coverage, such as mega-\ac{LEO} constellations, Shewhart-based access schemes prove advantageous by reducing the required bandwidth to transmit target effective data.
{\renewcommand{\arraystretch}{1.3}
\begin{table}[t]
\caption{Performance of various regression models for predicting PM2.5 using PM10 at the server}
\label{tab:prediction}
\resizebox{\columnwidth}{!}{%
\begin{tabular}{|l|l|l|l|}
\hline
\textbf{Model} & \textbf{Configuration} & \textbf{\begin{tabular}[c]{@{}l@{}}RMSE\\ (ppm)\end{tabular}} & \textbf{\begin{tabular}[c]{@{}l@{}}R-Squared\\ Validation\end{tabular}} \\ \hline \hline
Linear Regression & Ordinary least squares & 2.1297 & 0.9808 \\ \hline
Decision Tree Regressor & \begin{tabular}[c]{@{}l@{}}Max. depth = 10\\ Min samples split = 5\end{tabular} & 2.3587 & 0.9806 \\ \hline
Random Forest Regressor & \begin{tabular}[c]{@{}l@{}} No. of estimators = 100\\ Max. depth = 10\\ Min. samples split = 5\end{tabular} & 2.0563 & 0.9835 \\ \hline
\end{tabular}%
}
\end{table}
}
\subsection{Performance of ML algorithms for parameter prediction}
Three machine learning-based algorithms—linear regression, decision tree regression, and \ac{RF} regression were trained on the dataset to assess their effectiveness in predicting PM2.5 using PM10. Table \ref{tab:prediction} presents the average performance of these three schemes across the entire dataset, measured in terms of \ac{RMSE} and R-squared validation, with the selected configuration of the algorithms. The training was performed using the standard \textit{Sci-kit} learn implementation of these algorithms in Python. Given the heavy correlation in PM values, the performance of all three schemes is comparable. However, the \ac{RF} outperforms the others with an \ac{RMSE} of nearly 2 ppm, a negligible value when considering the nominal PM values in the winter season and the threshold set for Shewart-access schemes. While all three algorithms can be implemented on cloud servers in real-time, as validated in \cite{adarsh}, considering computational complexity becomes pivotal in making the final implementation choice. Regarding computational complexity, linear regression is the simplest among the three, involving only two weight parameters to learn.

Additionally, the \ac{ML} schemes have a pivotal role in reducing the size of the payload (volume of data) to be transmitted since the transmission of correlated parameters can be avoided without losing any significant information. The impact of reduced payload due to these \ac{ML} algorithms on the battery lifetime is analyzed in the subsequent subsection.
\begin{figure}[t!]
\centering
\includegraphics[width=0.95\columnwidth, keepaspectratio]{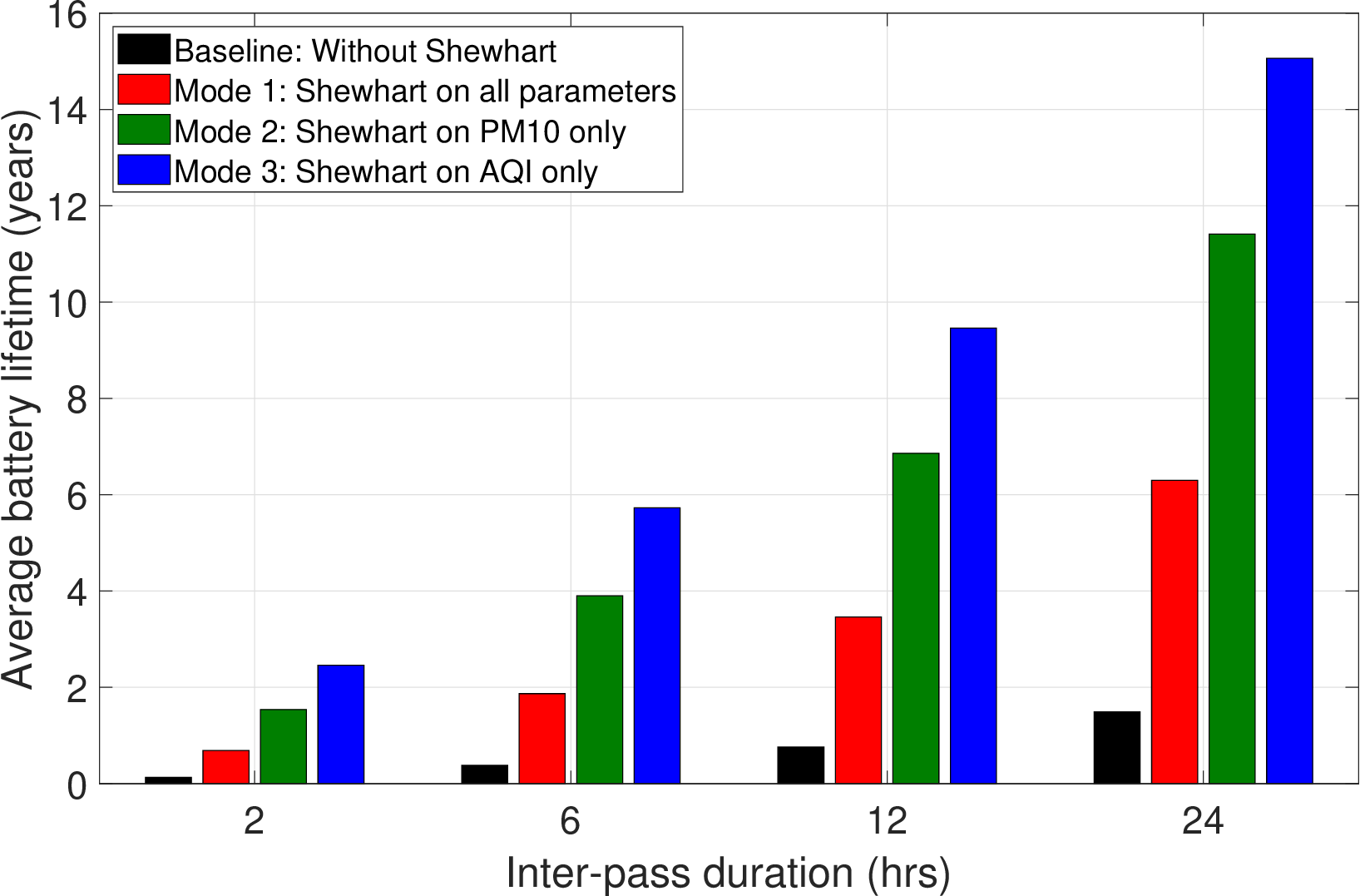}
\caption{Estimated lifetime of a 5000 mWh battery for 164 dB MCL and 5-minute visibility duration per pass for various modes of operation. The inter-pass duration represents the period between two consecutive instances when the satellite becomes visible and the IoT devices wake up for sensing and transmission. The devices remain in sleep mode when the satellite is not in the visible range.}
\label{fig:lifetime}
\end{figure}
\subsection{Expected battery life} \label{sec:batterylifetime}
Fig. \ref{fig:lifetime} shows the expected average battery lifetime of a 5000 mWh battery under specific conditions, such as a maximum coupling loss (MCL) of 164 dB (worst-case scenario as per \cite{standard_1}), 5-minute satellite visibility per pass, and varying inter-pass durations within a sparse satellite constellation. The inter-pass duration represents the period between two consecutive instances when the satellite becomes visible in a constellation with discontinuous coverage. The \ac{IoT} devices wake up for sensing and transmission only when the satellite is visible and remain in sleep mode otherwise. Following the recommendations made by \ac{3GPP} in \cite{standard_1}, assumptions made for calculations in this subsection include nodes possessing GNSS capabilities for accurate timing and frequency offset compensation, absence of simultaneous GNSS and \ac{NB-IoT} operations, no re-transmissions, and negligible power consumption in reading satellite ephemeris information. Table \ref{tab:power} shows the time and power required for various operations in \ac{NB-IoT} communication used for calculations in this section, as referred from \cite{standard_1}. Fig. \ref{fig:lifetime} highlights that modes incorporating Shewhart exhibit significantly prolonged battery lifetimes compared to the baseline mode without Shewhart. In this figure, Mode 1 has a payload of 200 bytes while Mode 2 and Mode 3 have a payload of 50 bytes, considering that only a few parameters are transmitted, and the others can be predicted using \ac{ML} algorithms on the server. The estimated enhancement in battery lifetime, coupled with the capacity to accommodate more devices, underscores the viability of the proposed Shewhart-based access scheme. Additionally, extending the inter-pass duration is observed to increase battery lifetime further, with improvements ranging from nearly two times to three times across different modes as the duration increases from 2 to 6 hours.

Table \ref{tab:batteryLifetime} shows the battery lifetime of all transmission modes, distinguishing between a full 200-byte payload and a reduced 50-byte payload. Notably, in Mode 2, reducing the payload leads to a 1.7-fold increase in battery life across all inter-pass durations, emphasizing the effectiveness of the proposed \ac{ML} algorithms in payload reduction. Similarly, in Mode 3, a 1.5-fold improvement in battery lifetime is observed with a reduced payload.
{\setlength{\tabcolsep}{3pt}
{\renewcommand{\arraystretch}{1.25}
\begin{table}[t!]
\centering
\caption{Power consumption assumptions for NTN-IoT energy consumption analysis \cite{standard_1}}
\label{tab:power}
\resizebox{\columnwidth}{!}{%
\begin{tabular}{|c|c|c|c|}
\hline
\textbf{State} & \textbf{Operation} & \textbf{\begin{tabular}[c]{@{}c@{}}Duration\\ (ms)\end{tabular}} & \textbf{\begin{tabular}[c]{@{}c@{}}Power\\ (mW)\end{tabular}} \\ \hline \hline
Reception (DL) & \begin{tabular}[c]{@{}c@{}}Sync, MIB, RAR Msg2, Msg4,\\ UL grant, HARQ ACK,\\ IP ACK, PDCCH monitoring,\end{tabular} & 371 & 90 \\ \hline
\multirow{2}{*}{Transmission (UL)} & \multirow{2}{*}{\begin{tabular}[c]{@{}c@{}}PRACH, RA Msg3 RAR,\\ IP Report, HARQ ACK\end{tabular}} & 50B UL: 335 & \multirow{2}{*}{543} \\ \cline{3-3}
 &  & 200B UL: 1006 &  \\ \hline
Idle (not in sleep) & \begin{tabular}[c]{@{}c@{}}MIB acquisition, waiting IP ACK,\\ PRACH, ready timer, scheduling\end{tabular} & 22423 & 2.4 \\ \hline
Power save (sleep) & \begin{tabular}[c]{@{}c@{}}Sleeping state when the satellite\\ is not visible\end{tabular} & \begin{tabular}[c]{@{}c@{}}based-on\\ visibility\end{tabular} & 0.015 \\ \hline
GNSS & GNSS reception & 2000 & 37 \\ \hline
\end{tabular}%
}
\end{table}
}}
%
%
%
% MCL = 154dB
% {\setlength{\tabcolsep}{3pt}
% {\renewcommand{\arraystretch}{1.3}
% \begin{table}[t!]
% \centering
% \caption{Average battery lifetime (in years) per device for MCL = 154 dB, 5 min visibility duration per pass and 5000 mWh battery}
% \label{tab:batteryLifetime}
% \resizebox{\columnwidth}{!}{%
% \begin{tabular}{|c|c|c|cc|cc|}
% \hline
% \multirow{3}{*}{\textbf{\begin{tabular}[c]{@{}c@{}}Inter-pass\\ duration\\ (hrs)\end{tabular}}} & \multirow{3}{*}{\textbf{Baseline}} & \multirow{3}{*}{\textbf{Mode 1}} & \multicolumn{2}{c|}{\textbf{Mode 2}} & \multicolumn{2}{c|}{\textbf{Mode 3}} \\ \cline{4-7} 
%  &  &  & \multicolumn{1}{c|}{\multirow{2}{*}{\textbf{200B UL}}} & \multirow{2}{*}{\textbf{50B UL}} & \multicolumn{1}{c|}{\multirow{2}{*}{\textbf{200B UL}}} & \multirow{2}{*}{\textbf{50B UL}} \\
%  &  &  & \multicolumn{1}{c|}{} &  & \multicolumn{1}{c|}{} &  \\ \hline \hline
% 2 & 0.56 & 3.18 & \multicolumn{1}{c|}{4.19} & 7.30 & \multicolumn{1}{c|}{8.30} & 13.17 \\ \hline
% 6 & 1.63 & 7.58 & \multicolumn{1}{c|}{8.92} & 14.16 & \multicolumn{1}{c|}{14.09} & 20.16 \\ \hline
% 12 & 3.13 & 12.24 & \multicolumn{1}{c|}{13.80} & 19.92 & \multicolumn{1}{c|}{19.23} & 25.25 \\ \hline
% 24 & 5.79 & 18.18 & \multicolumn{1}{c|}{19.74} & 25.65 & \multicolumn{1}{c|}{25.26} & 30.16 \\ \hline
% \end{tabular}%
% }
% \end{table}
% }}
%
% MCL = 164 dB
{\setlength{\tabcolsep}{3pt}
{\renewcommand{\arraystretch}{1.3}
\begin{table}[t!]
\centering
\caption{Average battery lifetime (in years) per device for MCL = 164 dB, 5 min visibility duration per pass and 5000 mWh battery}
\label{tab:batteryLifetime}
\resizebox{\columnwidth}{!}{%
\begin{tabular}{|c|c|c|cc|cc|}
\hline
\multirow{3}{*}{\textbf{\begin{tabular}[c]{@{}c@{}}Inter-pass\\ duration\\ (hrs)\end{tabular}}} & \multirow{3}{*}{\textbf{Baseline}} & \multirow{3}{*}{\textbf{Mode 1}} & \multicolumn{2}{c|}{\textbf{Mode 2}} & \multicolumn{2}{c|}{\textbf{Mode 3}} \\ \cline{4-7} 
 &  &  & \multicolumn{1}{c|}{\multirow{2}{*}{\textbf{200B UL}}} & \multirow{2}{*}{\textbf{50B UL}} & \multicolumn{1}{c|}{\multirow{2}{*}{\textbf{200B UL}}} & \multirow{2}{*}{\textbf{50B UL}} \\
 &  &  & \multicolumn{1}{c|}{} &  & \multicolumn{1}{c|}{} &  \\ \hline \hline
2 & 0.13 & 0.69 & \multicolumn{1}{c|}{0.87} & 1.54 & \multicolumn{1}{c|}{1.60} & 2.46 \\ \hline
6 & 0.38 & 1.87 & \multicolumn{1}{c|}{2.23} & 3.90 & \multicolumn{1}{c|}{3.67} & 5.73 \\ \hline
12 & 0.76 & 3.46 & \multicolumn{1}{c|}{4.04} & 6.86 & \multicolumn{1}{c|}{6.23} & 9.46 \\ \hline
24 & 1.49 & 6.30 & \multicolumn{1}{c|}{7.18} & 11.41 & \multicolumn{1}{c|}{10.62} & 15.06 \\ \hline
\end{tabular}%
}
\end{table}
}}
%
%------------Conclusion----------------
\section{Conclusion}
\label{sec:conclusion}
This paper introduced an efficient access scheme tailored for LEO satellite-based NB-IoT networks, employing a constellation with intermittent coverage and analysing data from a real-world IoT testbed. The analysis reveals distinctive traffic characteristics in the proposed Shewhart-based access scheme, deviating from the commonly assumed Poisson arrivals in random access scenarios. Notably, the Shewhart traffic significantly reduces the number of simultaneously transmitting nodes and associated collision probability compared to the baseline without transmission reduction. This translates to a significant increase in network capacity, accommodating more nodes while meeting collision probability targets. Despite the reduced transmissions, the Shewhart-based scheme delivers a significantly larger effective data volume at the server. Additionally, integrating machine learning algorithms for payload reduction substantially extends the battery lifetimes of the IoT devices. Consequently, the proposed Shewhart-based access scheme, combined with ML algorithms, emerges as a compelling solution for addressing the challenges of limited visibility, low data rates, and energy constraints in satellite-based IoT networks. In the future, the authors would like to develop an analytical model for the traffic pattern that best fits the empirical distribution and express it as a function of the sensed parameters. This would help make the change detection adaptable to the physical environment and further improve the performance.
\bibliographystyle{IEEEtran}
\bibliography{IEEEabrv,ref.bib}{}
\end{document}